\documentclass[floats,floatfix,showpacs,preprintnumbers,amssymb,prd,superscriptaddress,nofootinbib,nolongbibliography,reprint]{revtex4-1}
\usepackage[dvipdfmx]{graphicx}

\usepackage{subfigure}
\usepackage{bm}
\usepackage{mathrsfs}
\usepackage{amsmath}
\usepackage{cases}
\usepackage{multirow}

\def\be{\begin{equation}}
\def\ee{\end{equation}}
\newcommand{\beq}{\begin{eqnarray}}
\newcommand{\eeq}{\end{eqnarray}} 
\usepackage{color}
\usepackage{ulem}
\usepackage[linktocpage=true]{hyperref}

\begin{document}
\title{Theoretical modeling of approximate universality of tidally deformed neutron stars}

\author{Takuya Katagiri}
\affiliation{Center of Gravity, Niels Bohr Institute, Blegdamsvej 17, 2100 Copenhagen, Denmark}

\author{Gowtham Rishi Mukkamala}
\affiliation{Center of Gravity, Niels Bohr Institute, Blegdamsvej 17, 2100 Copenhagen, Denmark}

\author{Kent Yagi}
\affiliation{Department of Physics, University of Virginia, Charlottesville, Virginia 22904, USA}

\date{\today}

\begin{abstract}
Quasi-universal relations are known to exist among various neutron star observables that do not depend sensitively on the underlying nuclear matter equations of state.
For example, some of these relations imply that the tidally induced multipole moments are approximately characterized by the electric-type quadrupolar tidal deformability. 
Such relations can be used to reduce the number of independent tidal parameters in gravitational-waveform modeling, thereby allowing us to infer extreme nuclear matter properties more accurately and test General Relativity in an insensitive manner to uncertainties in nuclear physics. 
We present a comprehensive theoretical investigation into approximate universality of neutron stars. Our approach employs a semi-analytic relativistic stellar interior model, which extends the Tolman~VII solution, thereby enabling a refined exploration of the tidal properties of nonrotating stars within a semi-analytic framework. The derived power-law relations among various tidal deformabilities -- referred to as the universal Love relations -- agree well with  expressions in previous work found empirically. We elucidate how the equation-of-state dependence is suppressed in a particular combination of macroscopic physical parameters induced by perturbations and demonstrate that the relation between the moment of inertia and electric-type quadrupolar tidal deformability (I-Love relation) rests on the same underlying mechanism. Our findings indicate that the approximate universality of neutron stars can be attributed to low compressibility, consistent with some of the previous studies on the possible origin of the universality.

\end{abstract}

\maketitle

\section{Introduction and summary}

\subsection{Neutron star physics}
Neutron stars~(NSs) exhibit a variety of observable phenomena, including gravitational waves~(GWs), neutrino emissions, and electromagnetic waves in the broad frequency band. With their rich information channels, NSs serve as powerful probes for addressing outstanding questions in astrophysics, nuclear physics, and fundamental physics. The first detection of GWs from the binary NS merger, GW170817~\cite{LIGOScientific:2017vwq,LIGOScientific:2018hze}, was accompanied by electromagnetic counterparts, ushering us into the era of multi-messenger astronomy~\cite{LIGOScientific:2017ync}. This event provided evidence that binary NS mergers can be a source of heavy $r$-process elements~\cite{Kasen:2017sxr} and  main engines of short gamma-ray bursts~\cite{LIGOScientific:2017ync}.    

NSs are the densest objects in the Universe, offering a unique opportunity to explore the properties of matter at supranuclear densities~\cite{Lattimer:2015nhk,Ozel:2016oaf,Burgio:2021vgk}. The central density is a few times the nuclear saturation density~$\rho_{\rm nuc}=2.8\times 10^{14}~{\rm g~cm^{-3}}$, a regime that cannot yet be accessed from the first principles. Determining the equation of state~(EoS) of nuclear matter in an extreme environment is one of the outstanding goals in nuclear physics. The NS mass-radius relation is in one-to-one correspondence with the EoS. Therefore, independent accurate measurements of the NS mass and radius with multifaceted observations allow us to determine the EoS. NS masses have been accurately measured through the Shapiro time delay (and other post-Keplerian parameters) in the arrival times of radio pulses emitted by millisecond pulsars~\cite{Jacoby:2005qg,Verbiest:2008gy,Demorest:2010bx}. NS radii, together with masses, are determined by the NICER mission using X-ray that is emitted from hot spots on the surfaces of millisecond pulsars~\cite{Riley_2019,Miller_2019,Miller_2021,Choudhury_2024}. GW observations of binary NSs allow us to constrain the EoS and radii by measuring the stellar tidal deformability~\cite{LIGOScientific:2017vwq,LIGOScientific:2018cki,LIGOScientific:2018hze,LIGOScientific:2020aai}.

NSs, in the context of fundamental physics, serve as natural laboratories for strong-field tests of gravity~\cite{Will:2005va,Berti:2015itd,Cardoso:2017cfl,Kramer:2021jcw,Shao:2022koz,Yunes:2024lzm}. In the weak-field regime,\footnote{The weak-field regime is characterized by a regime where the ratio of the mass to the characteristic length scale of the system, as well as the characteristic velocity, are much smaller than unity in the geometric units. The strong-field regime, on the other hand, is characterized by a regime where the ratio is of the order of unity or less, and the characteristic velocity is much smaller than unity} the first observational experiment using NSs was unexpectedly conducted by measuring the decay rate of the orbital period of the Hulse-Taylor binary radio pulsar~\cite{Hulse:1974eb}, which is consistent with the theoretical predictions of GW emission in General Relativity~(GR), thereby providing empirical support for the theory. The compactness of NSs typically ranges from $0.1$ to $0.3$, and hence, their structure lies in the strong-field regime. Alternative theories of gravity may provide different descriptions of the coupling between gravity and matter from GR. Consequently, the mass-radius relation differs from that within GR for a given EoS, and in turn, their observational signature is also modified from the predictions by GR~\cite{Pani:2011xm,Pani:2014jra,Yagi:2021loe,Creci:2023cfx,vanGemeren:2023rhh,Creci:2024wfu,Diedrichs:2025vhv} (see also~\cite{Yagi:2013qpa,Yagi:2013ava,Yagi:2013mbt,Saffer:2021gak} for the NS mass-radius relation in non-GR theories that do not alter the matter-gravity coupling from GR but modify the field equations).  

The aforementioned properties of NS structures and observational outcomes suggest that NSs are ideal laboratories for studying fundamental interactions among physical degrees of freedom in extremely dense and strong-gravity environments, including not only known components, but also hypothetical sectors. However, there are uncertainties in both the EoS and the underlying gravitational theory in the NS interiors. The achievement of the determination of the fundamental interactions thus requires breaking the degeneracies within the two types of uncertainties (nuclear and gravitational physics).

\subsection{Approximate universality for neutron stars}
Spinning and tidally-deformed NSs exhibit approximate universality known as the ``I-Love-Q" relations, which link the moment of inertia~(I), tidal deformability~(Love), and spin induced mass quadrupole moment~(Q) in an EoS insensitive way~\cite{Yagi:2013bca,Yagi:2013awa,Yagi:2016bkt}.\footnote{Although many studies assume slow rotation in the sense that the dimensionless spin parameter~$\chi(:=|\vec{S}|/M^2)$ for the spin angular momentum~$\vec{S}$ and mass~$M$ is much smaller than unity, the I-Love relation remains valid for arbitrary rotation~\cite{Pappas:2013naa,Chakrabarti:2013tca,Yagi:2014bxa}. } For instance, the I-Q curve for NSs across various EoSs, including realistic EoS models proposed from nuclear physics, remains nearly indistinguishable, allowing us to approximately express ``I" as a function of ``Q" independently of EoSs, and vice versa. Consequently, once one of these parameters is determined, the others can be inferred, thereby breaking the degeneracies among the parameters used in the theoretical modeling of X-ray light curves from hot spots on NS surfaces~\cite{Baubock:2013gna}. The I-Love-Q relation provides an unprecedented avenue for determining NS parameters by combining independent observations with GWs and electromagnetic waves. A similar approximate universality extends to other stellar parameters. The multipole Love relations discovered in Ref.~\cite{Yagi:2013sva} connect tidal parameters at different post-Newtonian~(PN) orders in the GW phase in a manner insensitive to EoSs. Furthermore, for cold and old NSs with realistic EoSs, the I-Love-C relation for the stellar compactness~C is also insensitive to the EoSs within a relative error~$\lesssim 2\%$~\cite{Maselli:2013mva,Yagi:2016bkt}. Other extensions include the relation between the f-mode frequency and the tidal deformability~\cite{Chan:2014kua}~(further extensions involving the first $p$ and $w$ modes are studied in~\cite{Sotani:2021kiw}), three-hair relations among various multipole moments~\cite{Stein:2013ofa,Yagi:2014bxa}, for different stellar configurations with pressure anisotropy~\cite{Yagi:2015hda}, non-barotropic fluids~\cite{Martinon:2014uua}, differential rotation~\cite{Bretz:2015rna}, and magnetars~\cite{Haskell:2013vha,Zhu:2020imp}.

The universal Love relations, which refer to the empirical power-law relations found in Ref.~\cite{Yagi:2013sva}, connect approximately (conservative) tidal deformabilities in an EoS insensitive manner, thereby reducing the number of independent parameters in GW data analysis. With the third-generation detector~\cite{Punturo:2010zz}, neglecting the magnetic-type quadrupolar tidal deformability in the waveform template may lead to a small bias in the parameter estimation~\cite{JimenezForteza:2018rwr}. Given recent advance of waveform modeling up to higher PN-order tidal corrections~\cite{Yagi:2013sva,Abdelsalhin:2018reg,Banihashemi:2018xfb,JimenezForteza:2018rwr,Landry:2018bil,Henry:2020ski,Castro:2022mpw,Narikawa:2023deu}, it is beneficial to investigate these empirical relations in greater depth by rederiving the theoretical expressions in a (semi-)analytic manner.

It is natural to ask what is the physical origin of the approximate universality found in NSs. At present, there is no decisive answer, although some evidence exist -- such as the approximate self-similarity of isodensity contours in stellar interiors~\cite{Yagi:2014qua}, the incompressible limit as a stationary point of the relative change rate of the I-Love relation for different EoSs~\cite{Sham:2014kea}, and the monotonic behavior exhibited by certain perturbation variables in their radial profiles~\cite{Kyutoku:2025zud}.  A deeper exploration of the underlying structure of the universality will illuminate unexplored aspects of how stellar characters -- originally governed by a number of parameters -- tend to be lost as stellar configurations evolve along the sequence from non-relatistic to relativistic stars in a phase space~\cite{Yagi:2014qua}, where the extreme limit is a black hole with the no-hair property.

\subsection{Summary of this work}

\begin{widetext}
    
\begin{figure}[htbp]
\begin{minipage}[b]{0.3 \linewidth}
  \centering
  \hspace{-4cm}
 \includegraphics[scale=0.73]{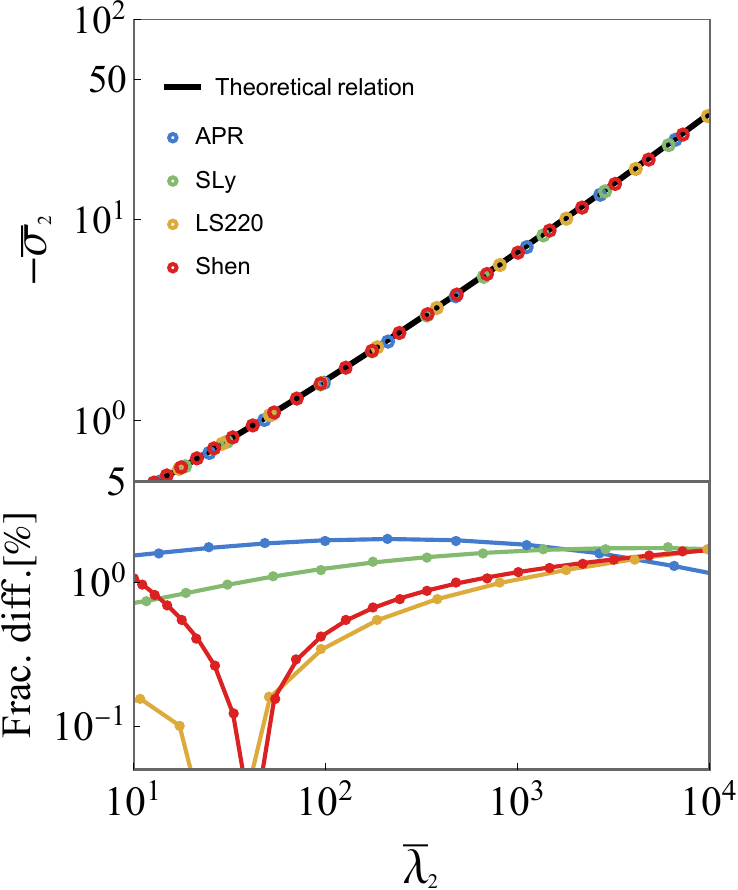}
 \end{minipage}
 \begin{minipage}[b]{0.3 \linewidth}
  \centering
   \hspace{5cm}
  \includegraphics[scale=0.70]{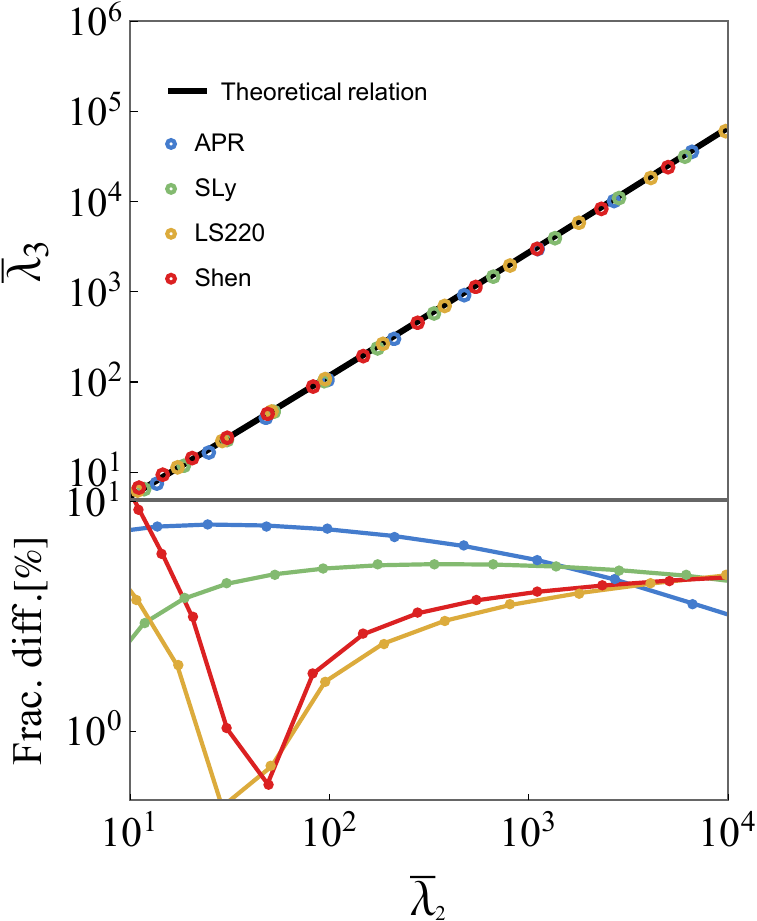}
 \end{minipage}
\caption{(Top)~Approximately universal $\bar{\lambda}_2-\bar{\sigma}_2$ (left) and $\bar{\lambda}_2-\bar{\lambda}_3$ (right) relations. The black lines in the left and right panels are the power-law relations~\eqref{eq:lambda2sigma2} and~\eqref{eq:lambda2lambda3}, respectively. The colored circles denote a set of $(\bar{\lambda}_2,\bar{\sigma}_2)$ and $(\bar{\lambda}_2,\bar{\lambda}_3)$ that were computed numerically for realistic EoS models~(APR~\cite{Akmal:1998cf}, SLy~\cite{Douchin:2001sv}, LS220~\cite{Lattimer:1991nc}, and Shen~\cite{Shen:1998gq,Shen:1998by}). Observe that the relations with different EoSs are indistinguishable. (Bottom)~ The fractional difference of the relations for realistic EoSs from the power-law expressions~\eqref{eq:lambda2sigma2} and~\eqref{eq:lambda2lambda3}.}
\label{Fig:UniversalLove}
\end{figure}
\end{widetext}

We conduct a comprehensive, theoretical study on the approximate universality of NSs. Our approach employs a semi-analytic relativistic stellar interior model, referred to as a modified Tolman~VII model~\cite{Posada:2022lij,Jiang:2019vmf}, and solves the tidal perturbations based on the post-Minkowskian expansion. We semi-analytically derive the universal Love relations among the tidal deformabilities of nonrotating NSs by solving tidal perturbations on the Tolman~VII solution~\cite{PhysRev.55.364}, which is a particular case of the modified Tolman~VII model:
\begin{widetext}
\begin{align}
    \bar{\sigma}_2= -0.0135394 \bar{\lambda}_2^{4/5} \left(1+\frac{3.06615}{\bar{\lambda}_2^{1/5}}+\frac{3.69338}{\bar{\lambda}_2^{2/5}}  +\frac{1.98439}{\bar{\lambda}_2^{3/5}} +\frac{0.343307}{\bar{\lambda}_2^{4/5}} +\frac{0.0875811}{\bar{\lambda}_2}\right),\label{eq:lambda2sigma2}
\end{align}
and
\begin{align}
    \bar{\lambda}_3=0.155283 \bar{\lambda}_2^{7/5} \left(1+ \frac{0.342365}{\bar{\lambda}_2^{1/5}}+\frac{0.300941}{\bar{\lambda}_2^{2/5}} +\frac{0.085495}{\bar{\lambda}_2^{3/5}} +\frac{0.142828}{\bar{\lambda}_2^{4/5}}+\frac{0.0435432}{\bar{\lambda}_2}  \right),\label{eq:lambda2lambda3}
\end{align}
\end{widetext}
where $\bar{\lambda}_\ell$ and $\bar{\sigma}_\ell$ are the $\ell$th electric-type and magnetic-type tidal deformabilities, respectively.\footnote{Note that $\bar{\lambda}_\ell$ and $\bar{\sigma}_\ell$ are dimensionless quantities and are identical to coefficients~$\Lambda_\ell$ and $\Sigma_\ell$ commonly used in the context of GW astronomy~\cite{Favata:2013rwa,Henry:2020ski,Narikawa:2023deu}.} These relations agree well with the empirical expressions in Ref.~\cite{Yagi:2013sva} that were obtained by fitting against numerical data. Figure~\ref{Fig:UniversalLove} presents the $\bar{\lambda}_2-\bar{\sigma}_2$ and $\bar{\lambda}_2-\bar{\lambda}_3$ relations with the power-law expressions in Eqs.~\eqref{eq:lambda2sigma2} and~\eqref{eq:lambda2lambda3}, demonstrating the approximate universality for tidally deformed NSs with realistic EoSs. Following Ref.~\cite{Yagi:2013sva}, we consider the following EoSs: Akmal, Pandharipande, and Ravenhall~(APR)~\cite{Akmal:1998cf}, Skyrme-Lyon~(SLy)~\cite{Douchin:2001sv},  Lattimer-Swesty with nuclear incompressibility of 220MeV~(LS220)~\cite{Lattimer:1991nc}, and Shen~\cite{Shen:1998gq,Shen:1998by}.

The tidally induced multipole moments of NSs are approximately characterized by $\bar{\lambda}_2$. Consequently, higher-PN-order tidal deformabilities in waveform models can be analytically expressed in terms of~$\bar{\lambda}_2$, reducing the number of independent parameters in GW data analysis, thereby improving the measurement accuracy of the leading tidal parameter. Furthermore, these relations impose theoretical constraints among the tidal deformabilities of NSs within GR, allowing us for strong-field tests of gravity in a manner insensitive to uncertainties in the EoSs.

We reveal the suppression mechanism of the EoS dependence in the universal relations in a particular combination of tidal Love numbers~(TLNs). This mechanism essentially relies on the macroscopic property of tidally deformed NSs: stiffer EoSs lead to stronger tidal deformation. The I-Love relation is based on a similar suppression mechanism between the moment of inertia and the TLN, in which stiffer EoSs lead to a higher moment of inertia. The shared EoS dependence among linear responses results in suppression of the EoS dependence. We speculate that NSs perturbed by arbitrary processes exhibit, more or less, a similar suppression mechanism in particular combinations among macroscopic physical parameters induced by perturbations.

The suppression mechanism is continuously reinforced as the local adiabatic index increases. Our findings suggest that the approximate universality of perturbed NSs can be attributed to low compressibility. Indeed, we find that stellar configurations in equilibrium tend to become insensitive to variations in EoSs as the local adiabatic index increases. These observations suggest the following picture: incompressible stars correspond to a restricted subset of equilibrium stellar configurations in the phase space spanned by physical degrees of freedom, as the energy density remains constant throughout the entire interior. Linear perturbations to such incompressible stars induce the strongest possible responses, setting upper bounds on their magnitudes. Stars deviate from this limit by decreasing the adiabatic index. For NSs, stellar configurations remain insensitive to variations in the EoSs due to low compressibility. A decrease in the adiabatic index reduces the magnitudes of the responses in a manner shared among the perturbations, leading to approximate universality. Eventually, as the adiabatic index decreases further, the degrees of freedom associated with an EoS expand, and in turn, the similarity of the responses is lost, resulting in the loss of universality.

\subsection{Organization and convention}
In Section~\ref{Section:ModifiedTolmanVII}, we introduce a semi-analytic model for relativistic stellar interiors. In Section~\ref{Sec:ApproximateUbniversalRelations}, we present the analytic, approximate universal relations among the tidal deformabilities of NSs. In Section~\ref{Section:PotentialOrigin}, we discuss the potential origin of these universal Love relations and unveil how EoS dependence is suppressed in a particular combination of TLNs. Section~\ref{Section:Discussion} is devoted to discussing our findings and mentioning possible future extensions. Appendix~\ref{Appendix:modifiedTolmanVII} introduces semi-analytic and fully analytic NS interior models used in this work. In Appendix~\ref{Appendix:LinearStellarPerturbation}, we summarize the linear perturbation theory of relativistic spherical stars within a perfect-fluid approximation. In Appendix~\ref{Appendix:ILove}, we demonstrate that the I-Love relation results from the same suppression mechanism of the EoS dependence as the universal Love relations. We adopt the geometric units with $c=G=1$.

\section{Semi-analytic model of neutron star interior}\label{Section:ModifiedTolmanVII}
We consider a static and spherically symmetric relativistic star in isolation. In static coordinates~$(t,r,\theta,\varphi)$, the line element takes the form,
\begin{align}
    g_{\mu \nu} dx^\mu dx^\nu=-e^{\nu}dt^2+e^{\lambda} dr^2+r^2d\Omega^2,
\end{align}
where $\nu=\nu(r)$, $\lambda=\lambda(r)$, and $d\Omega^2:= d\theta^2+\sin^2\theta d\varphi^2$. We model a NS by a sphere occupied by an isotropic fluid in static equilibrium. The star should satisfy Einstein's equations,
\begin{align}
    G_{\mu\nu}=8\pi T_{\mu\nu},\label{eq:EinsteinEqscov}
\end{align}
with the Einstein tensor~$G_{\mu\nu}$ and the stress-energy tensor of a perfect fluid,
\begin{align}
    T_{\mu\nu}= \left(p+\rho\right)u_\mu  u_\nu +pg_{\mu\nu}.
\end{align}
Here, $p=p(r)$ and $\rho=\rho(r)$ are the pressure and energy density, respectively. The vector~$u^\mu$ is the four-velocity of the fluid, normalized by $g_{\mu\nu}u^\mu u^\nu=-1$, and hence, $u^\mu=(e^{-\nu/2},0,0,0)$. 

Einstein's equations~\eqref{eq:EinsteinEqscov} reduce to a set of the following three equations:
\begin{align}
    m'=&4\pi r^2\rho,~~\nu'=2\frac{m+4\pi r^3 p}{r\left(r-2m\right)},\nonumber\\
    p'=&-\left(p+\rho\right)\frac{m+4\pi r^3 p}{r\left(r-2m\right)},~\label{eq:EinsteinEqs}
\end{align}
where the prime denotes a differentiation with respect to $r$. Here, we have introduced a radial mass function~$m=m(r)$ defined by $m:=(r/2)(1-e^{-\lambda})$. To close Eq.~\eqref{eq:EinsteinEqs}, an EoS $p=p(\rho)$ connecting the internal matter degrees of freedom is necessary. The exterior solution to Eq.~\eqref{eq:EinsteinEqs}, where $\rho=p=0$, must be the Schwarzschild metric by virtue of Birkoff's theorem,
\begin{align}
    e^\nu=e^{-\lambda}=1-\frac{2M}{r},\label{eq:Schwarzschildmetric}
\end{align}
where $M$ is the gravitational mass.
In general, solving Eq.~\eqref{eq:EinsteinEqs} requires numerical calculations. 

Since this work aims to understand the tidal property of NSs semi-analytically, we follow the idea of a (semi-)analytic model for NS interiors, an extension of a Tolman~VII solution~\cite{PhysRev.55.364} by introducing an additional parameter~$\alpha$, proposed in Refs.~\cite{Jiang:2019vmf}. The parameter~$\alpha$ controls the internal relation between energy density and pressure, and hence, corresponds to the degree of freedom of an EoS. The energy density profile is given by
\begin{align}
    \rho=&\rho_c\left[1-\alpha \xi^2+\left(\alpha-1\right)\xi^4\right].\label{eq:densityofmTolman}
\end{align}
Here, $\rho_c$ is the central energy density and is related to $M$ and $\alpha$~(see Eq.~\eqref{eq:rhoc}); $\xi:=r/R$ is a dimensionless radial coordinate normalized by the radius~$R$. Given $\rho$ in Eq.~\eqref{eq:densityofmTolman}, one can determine~$m$ fully analytically, while $\nu$ and $p$ are obtained by solving Eq.~\eqref{eq:EinsteinEqs} numerically, thereby constructing a semi-analytic NS interior model~\cite{Posada:2022lij}. The functions,~$m$, $\nu$, and $p$, are provided in Appendix~\ref{Appendix:modifiedTolmanVII}. We refer to this semi-analytic solution as a modified Tolman~VII solution. The fully analytic model proposed by Ref.~\cite{Jiang:2019vmf} provides a good approximation to this numerical solution, except for $p$ in the outer layers~\cite{Posada:2022lij}, despite not being an exact solution to Eq.~\eqref{eq:EinsteinEqs}. We provide the expressions for $m$, $\nu$, and $p$ of the fully analytic model in Appendix~\ref{Appendix:modifiedTolmanVII}.

The modified Tolman~VII solution has three independent parameters, $M$, $R$, and $\alpha$. The stellar configuration for $0.4 \lesssim \alpha \lesssim 1.4$ can model NSs well with various realistic EoSs in a wide range of compactness~\cite{Jiang:2019vmf,Jiang:2020uvb}. In this work, we assume $0\le \alpha \le 1.4$, as our purpose involves obtaining insights into the potential origin of the approximate universality for NSs. Taking $\alpha\to 1$, the modified Tolman~VII solution reduces to the Tolman~VII solution, originally derived in Ref.~\cite{PhysRev.55.364}. The quadratic energy density profile is a good approximation for stellar models with various realistic EoSs in a large portion of the interior~(see, e.g., Fig.~4 in Ref.~\cite{Sham:2014kea} and Fig.~9 in Ref.~\cite{Jiang:2019vmf}).

\begin{figure}[htbp]
\centering
\includegraphics[scale=0.63]{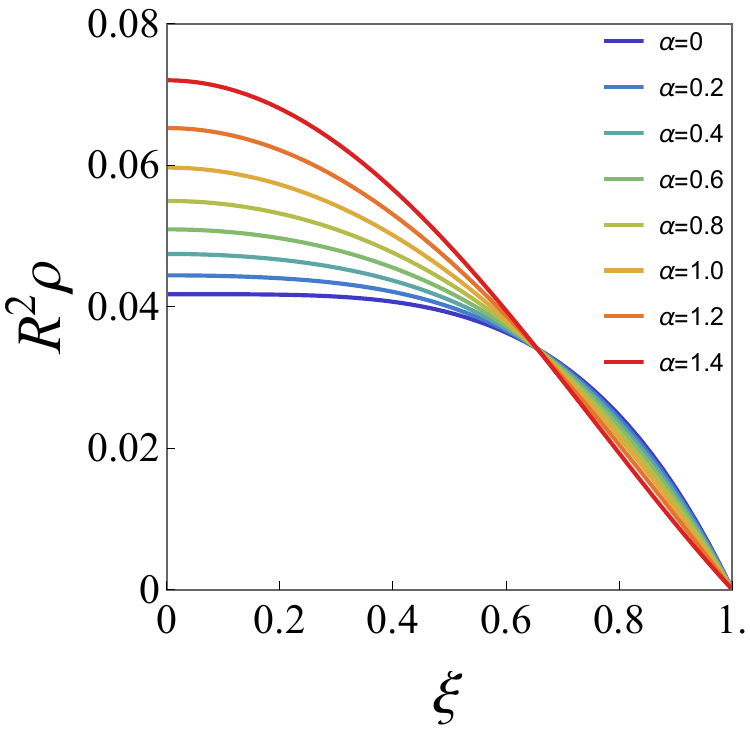}
\hspace{0.3cm}
\caption{The energy density profile in Eq.~\eqref{eq:densityofmTolman} for various~$\alpha$ with ${\cal C}=0.1$. Observe that the plateau near the stellar origin extends further as $\alpha$ decreases.
}
\label{Fig:EnergyDensityProfile}
\end{figure}

\begin{figure}[htbp]
\centering
\includegraphics[scale=0.60]{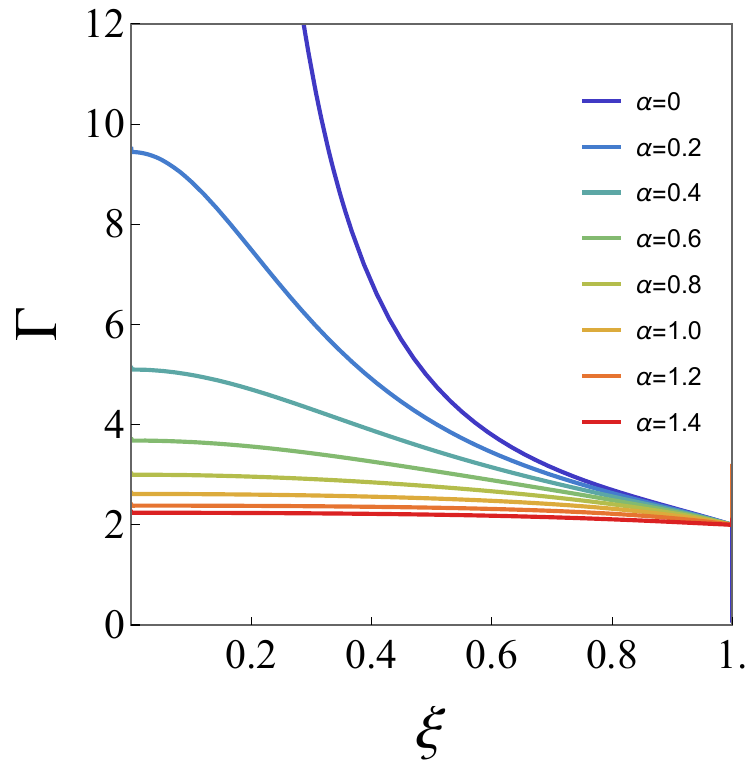}
\hspace{0.3cm}
\caption{The local adiabatic index~\eqref{eq:Gamma} for ${\cal C}=0.1$ with various~$\alpha$. Observe that smaller~$\alpha$ yields larger $\Gamma$. This behavior is qualitatively independent of the choice of ${\cal C}$. The blue curve~($\alpha=0$) diverges at the origin, being locally incompressible.
}
\label{Fig:Gamma}
\end{figure}

To see the role of $\alpha$, we briefly discuss the energy density profile~\eqref{eq:densityofmTolman}, presented in Fig.~\ref{Fig:EnergyDensityProfile}. Notice that $\rho_c$ tends to be lower with decreasing $\alpha$ and fixed compactness~${\cal C}(:=M/R)$.  A plateau exists near the stellar origin and extends as $\alpha$ decreases, suggesting that smaller~$\alpha$ corresponds to a stiffer EoS. The local stiffness of an EoS can be measured through the local adiabatic index defined by
\begin{align}
    \Gamma:=\left(1+\frac{p}{\rho}\right) \frac{d\ln p}{d\ln \rho}.\label{eq:Gamma}
\end{align}
Figure~\ref{Fig:Gamma} shows that smaller~$\alpha$ leads to larger~$\Gamma$. Note that the curves with different~$\alpha$ do not intersect. This implies that smaller~$\alpha$ corresponds to stiffer EoSs. In the limit of $\alpha\to 0$, $\Gamma$ blows up in the inner core, corresponding to a locally incompressible limit. 

Now, the dependence of $\rho$ on $\alpha$ in Fig.~\ref{Fig:EnergyDensityProfile} can be interpreted from a physical point of view. Smaller~$\alpha$ leads to lower compressible configurations in the entire interior, meaning that matter resists gravitational attraction more effectively, even at lower densities. Consequently, a lower central density is sufficient to generate the necessary pressure gradient to maintain equilibrium.

\section{Universal Love relations}\label{Sec:ApproximateUbniversalRelations}

\subsection{Tidal deformability}
Let us consider a star, initially spherically symmetric, immersed in a tidal environment created by other gravitational sources, such as a binary companion. The external source exerts tidal forces on the star, and in turn, the gravitational potential of the star receives anisotropic corrections, inducing higher-order multipole moments in its multipole expansion. If the external force is sufficiently weak and varies slowly in time compared to the timescale for the star settling down to equilibrium, the tidal interaction can be regarded as static tides, and thus, the analysis boils down to a stationary linear response problem for the star within the framework of relativistic perturbation theory.
 
The tidal perturbation is decomposed into the electric-type and magnetic-type sectors, subject to the parity transformation~\cite{Hinderer:2007mb,Damour:2009vw,Binnington:2009bb}. With harmonic decomposition, the tidal deformation of the star is quantified by a set of the dimensionless $\ell$th electric-type and magnetic-type tidal deformabilities $\bar{\lambda}_\ell$ and $\bar{\sigma}_\ell$~\cite{Damour:2009vw,Yagi:2013sva},
\begin{align}
    \bar{\lambda}_\ell: =& \frac{2}{\left(2\ell-1\right)!!} {\cal C}^{-2\ell-1}k_\ell,\label{eq:electriclambda}\\
     \bar{\sigma}_\ell : =& \frac{\ell-1}{4\left(\ell+2\right)\left(2\ell-1\right)!!}{\cal C}^{-2\ell-1}j_\ell.\label{eq:magneticsigma} 
\end{align}
Here, $k_\ell$ and $j_\ell$ are the $\ell$th electric-type and magnetic-type (dimensionless) TLNs, respectively.\footnote{This definition differs from that in Refs.~\cite{Katagiri:2023umb,Katagiri:2023yzm,Katagiri:2024fpn,Katagiri:2024wbg} by
\begin{align}
    j_\ell = \frac{2\left(\ell+2\right)\left(\ell+1\right)}{\ell\left(\ell-1\right)} \kappa_\ell^-.
\end{align}
Here, $j_\ell$ and $\kappa_\ell^-$ are the magnetic-type TLNs in the current work and in the literature, respectively.
} The TLNs can be read off from the large-distance expansion of the tidally deformed metric. Note that $\bar{\lambda}_\ell$ and $\bar{\sigma}_\ell$ are dimensionless quantities and are identical to coefficients~$\Lambda_\ell$ and $\Sigma_\ell$ commonly used in the context of GW astronomy~\cite{Favata:2013rwa,Henry:2020ski,Narikawa:2023deu}.\footnote{The tidal-polarizability coefficients~$\mu_\ell$ and $\sigma_\ell$ in Refs.~\cite{Damour:2009vw,Damour:2012yf} are of $[{\rm length}]^{2\ell+1}$ in the geometrical units and are related by $\mu_\ell= M^{2\ell+1}\bar{\lambda}_\ell$ and $\sigma_\ell=M^{2\ell+1} \bar{\sigma}_\ell$ for the stellar mass $M$.} The coefficients~$\bar{\lambda}_\ell$ and $\bar{\sigma}_\ell$ directly enter in the GW phase at $(2\ell+1)$ and $3\ell$ PN order, respectively, relative to the point-particle approximation~\cite{Flanagan:2007ix,Yagi:2013sva}.

\subsection{Power-law relations}
We here provide the analytic power-law expressions representing approximate universal relations among the tidal deformabilities. The perturbation equations to determine $\bar{\lambda}_\ell$ and $\bar{\sigma}_\ell$ are derived in Appendix~\ref{Appendix:LinearStellarPerturbation}. In Section~\ref{Section:PotentialOrigin}, we discuss the underlying structure of these relations in detail.

\subsubsection{Even--Odd Love relations}

\begin{figure}[htbp]
\centering
\includegraphics[scale=0.65]{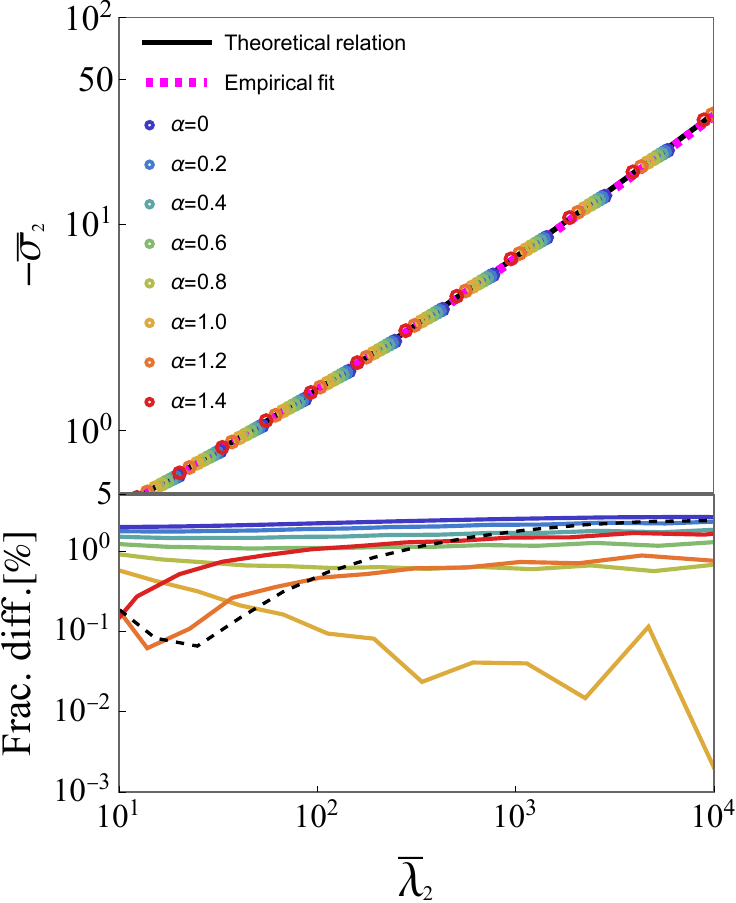}
\hspace{0.3cm}
\caption{(Top panel)~The~$\bar{\lambda}_2-\bar{\sigma}_2$ relation. The black line is the power-law relation~\eqref{eq:lambda2sigma2}. The magenta dashed line is the empirical fit for realistic EoSs in~\cite{Yagi:2013sva}. The colored circles denote a set of $(\bar{\lambda}_2,\bar{\sigma}_2)$ numerically computed for various~$\alpha$ in the modified Tolman VII model without relying on the PM expansion, given discrete compactness. Observe that the variation of the results with respect to $\alpha$ is indistinguishable. (Bottom panel)~ The fractional difference between the numerical values represented by the circles and the theoretical power-law relation~\eqref{eq:lambda2sigma2}~(solid curves). The black dashed curve denotes the fractional difference between the theoretical and fitting expressions. Note that the theoretical relation is approximately derived for the Tolman~VII solution~($\alpha=1$) within the PM expansion, leading to the smallest fractional difference in a wide range of $\bar{\lambda}_2$.}
\label{Fig:Lambda2Sigma2}
\end{figure}

Figure~\ref{Fig:Lambda2Sigma2} presents the $\bar{\lambda}_2 - \bar{\sigma}_2$ relation, demonstrating the insensitivity of the results with respect to variation in $\alpha$ and the excellent agreement of the power-law expression~\eqref{eq:lambda2sigma2} with the previous empirical finding by Ref.~\cite{Yagi:2013sva}.\footnote{Note that the bottom panel does not display the relative error between the semi-analytically constructed curves across various~$\alpha$ and the theoretical relation based on $\alpha=1$~(see the caption in Fig.~\ref{Fig:Lambda2Sigma2}). The finite difference for $\alpha=1$ merely reflects the accuracy of the PM expansion in that case. The same remark applies to Figs.~\ref{Fig:Lambda2Lambda3},~\ref{Fig:Lambda2num},~\ref{Fig:Sigma2num}, and~\ref{Fig:ILove}. } Note that the colored circles denote a set of $(\bar{\lambda}_2,\bar{\sigma}_2)$ numerically computed for various~$\alpha$ on the background described by the modified Tolman~VII solution.

The theoretical relation~\eqref{eq:lambda2sigma2} is derived in the following manner. First, we consider tidal perturbations in the fully analytic model of the modified Tolman~VII solution, provided in Appendix~\ref{Appendix:modifiedTolmanVII}. One then obtains the analytic expressions for $\bar{\lambda}_\ell$ and $\bar{\sigma}_\ell$ as a function of ${\cal C}$ for fixed~$\alpha$ by solving tidal perturbation equations within the post-Minkowskian~(PM) expansion, i.e., an expansion in terms of ${\cal C}$. The technical details are provided in Appendix~\ref{Appendix:LinearStellarPerturbation}.  The tidal deformabilities, $\bar{\lambda}_\ell$ and $\bar{\sigma}_\ell$, are expressed in the form of polynomials of ${\cal C}$. We re-sum those with the Pad{\'e} approximant to improve their accuracy, thereby obtaining
\begin{align}
    \bar{\lambda}_\ell= \frac{2}{\left(2\ell-1\right)!!} {\cal C}^{-2\ell-1}\frac{p_{\ell,0}^+ + p_{\ell,1}^+ {\cal C} + p_{\ell,2}^+ {\cal C}^2 + p_{\ell,3}^+ {\cal C}^3 }{ q_{\ell,0}^+ + q_{\ell,1}^+ {\cal C} + q_{\ell,2}^+ {\cal C}^2+ q_{\ell,3}^+ {\cal C}^3},\label{eq:barlambdainC}
\end{align}
and
\begin{align}
    \bar{\sigma}_\ell = \frac{\ell-1}{4\left(\ell+2\right)\left(2\ell-1\right)!!} {\cal C}^{-2\ell}\frac{p_{\ell,0}^- + p_{\ell,1}^- {\cal C} + p_{\ell,2}^- {\cal C}^2  }{ q_{\ell,0}^- + q_{\ell,1}^- {\cal C} + q_{\ell,2}^- {\cal C}^2 }, \label{eq:barsigmainC}
\end{align}
where $p_{\ell,i}^\pm =p_{\ell,i}^\pm(\alpha)$ and $q_{\ell,i}^\pm =q_{\ell,i}^\pm (\alpha)$. The explicit forms of $p_{\ell,i}^\pm$ and $q_{\ell,i}^\pm$ are provided in online~\cite{git}. Equations~\eqref{eq:barlambdainC} and~\eqref{eq:barsigmainC} agree with values computed numerically in a wide range of compactness, as presented in Appendix~\ref{Appendix:LinearStellarPerturbation}~(see Figs.~\ref{Fig:Lambda2num} and~\ref{Fig:Sigma2num}).

We now derive the power-law expression~\eqref{eq:lambda2sigma2}. Since the dependence of the $\bar{\lambda}_2-\bar{\sigma}_2$ relation on $\alpha$ is weak, we fix $\alpha=1$ without losing much generality. This choice corresponds to specifying the background using the original Tolman~VII solution~\cite{PhysRev.55.364}, allowing us to derive the power-law relation in a fully analytic manner consistent with Einstein's equations~\eqref{eq:EinsteinEqs}. Setting $\alpha=1$ and $\ell=2$, and solving Eq.~\eqref{eq:barlambdainC} about ${\cal C}$ inversely, we express ${\cal C}$ as a polynomial of $1/\bar{\lambda}_2^{1/5}$:
\begin{align}
    {\cal C}= \sum_{j=1}^8 \frac{\beta_j}{\bar{\lambda}_2^{j/5}},\label{eq:Cinbarlambda}
\end{align}
where $\beta_j$ are numbers given in Eq.~\eqref{eq:CoefficientsofC}. Substituting this into Eq.~\eqref{eq:barsigmainC} with $\ell=2$ and expanding it about $1/\bar{\lambda}_2^{1/5}$, one can express $\bar{\sigma}_2$ in the form of a power-law series of $\bar{\lambda}_2$, which is Eq.~\eqref{eq:lambda2sigma2}.

We now assess the $\alpha$ dependence of the power-law relation. Again, expressing ${\cal C}$ in terms of $\bar{\lambda}_2$ without fixing~$\alpha$, substituting this into Eq.~\eqref{eq:barsigmainC} with $\ell=2$, and expanding it about $1/\bar{\lambda}_2^{1/5}$, we obtain
\begin{align}
    \bar{\sigma}_2= -\bar{\lambda}_2^{4/5} \left[C_0^{\bar{\sigma}_2 \bar{\lambda}_2} +\frac{C_1^{\bar{\sigma}_2 \bar{\lambda}_2} }{\bar{\lambda}_2^{1/5}} +{\cal O}\left(\frac{1}{\bar{\lambda}_2^{2/5}}\right) \right],\label{eq:AnalyticEvenOddLove}
\end{align}
where $C_0^{\bar{\sigma}_2 \bar{\lambda}_2}=C_0^{\bar{\sigma}_2 \bar{\lambda}_2}\left(\alpha\right)$ and $C_1^{\bar{\sigma}_2 \bar{\lambda}_2}=C_1^{\bar{\sigma}_2 
\bar{\lambda}_2}\left(\alpha\right)$. The approximate universality arises from the insensitivity of the power-law coefficients to the $\alpha$-variation. Indeed, the Taylor expansion of $C_0^{\bar{\sigma}_2 \bar{\lambda}_2}$ and $C_1^{\bar{\sigma}_2 \bar{\lambda}_2}$ around $\alpha=1$ shows
\begin{align}
    C_0^{\bar{\sigma}_2 \bar{\lambda}_2}=& 0.014\left[1 + 0.028 \left(\alpha-1\right)+{\cal O}\left(\left(\alpha-1\right)^2\right)\right],\nonumber\\
     C_1^{\bar{\sigma}_2 \bar{\lambda}_2}=& 0.041\left[1 + 0.038 \left(\alpha-1\right)+{\cal O}\left(\left(\alpha-1\right)^2\right)\right]. 
\end{align}
Notice that the coefficient in front of $(\alpha-1)$ is less than $4\%$ of the leading-order one, showing the weak~$\alpha$ dependence. The universality is assessed by
\begin{align}
    \frac{\left. C_0^{\bar{\sigma}_2 \bar{\lambda}_2} \right|_{\alpha=0}}{\left. C_0^{\bar{\sigma}_2 \bar{\lambda}_2}\right|_{\alpha = 1.4}}\simeq 0.96,\label{eq:RatioinEvenOdd}
\end{align}
which indicates that the deviation of the $\bar{\lambda}_2-\bar{\sigma}_2$ relation in the range of $\alpha$ under consideration is only $4\%$. This is much smaller than the deviation between polytropes with $0\le n \le 1$, which was found to be~$11\%$~\cite{Yagi:2013sva}.

\subsubsection{Multipole Love relations}
\begin{figure}[htbp]
\centering
\includegraphics[scale=0.70]{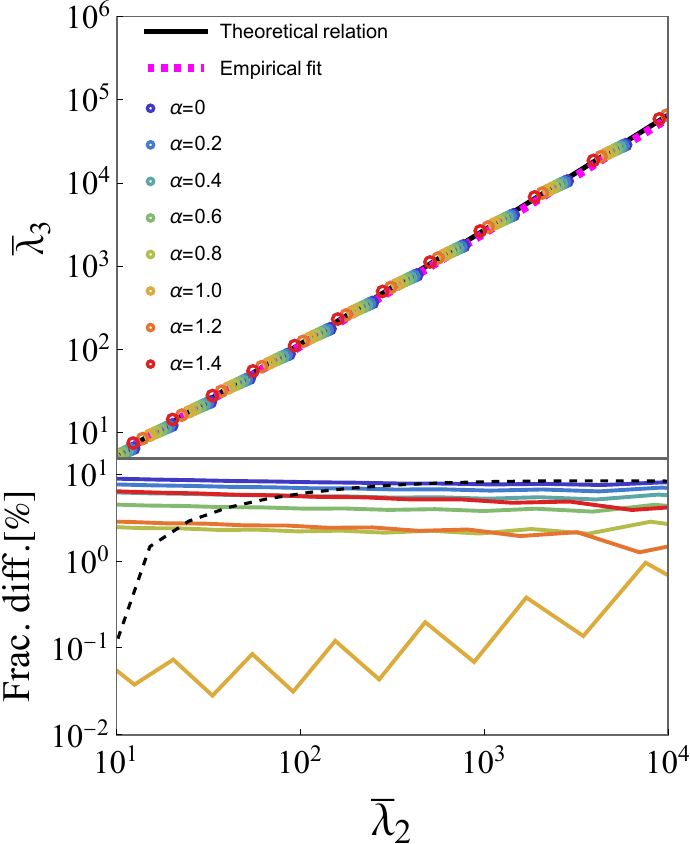}
\hspace{0.3cm}
\caption{Similar to Fig.~\ref{Fig:Lambda2Sigma2} but for the $\bar{\lambda}_2-\bar{\lambda}_3$ relation.}
\label{Fig:Lambda2Lambda3}
\end{figure}

Figure~\ref{Fig:Lambda2Lambda3} presents the $\bar{\lambda}_2-\bar{\lambda}_3$ relation, showing the weak~$\alpha$ dependence and agreement of the power-law relation~\eqref{eq:lambda2lambda3} with the fit found in Ref.~\cite{Yagi:2013sva}.  Notice that the colored circles denote a set of $(\bar{\lambda}_2,\bar{\lambda}_3)$ numerically computed for various~$\alpha$ on the modified Tolman~VII background. The theoretical expression~\eqref{eq:lambda2lambda3} is derived by substituting Eq.~\eqref{eq:Cinbarlambda} into Eq.~\eqref{eq:barlambdainC} with $\ell=2$ and $\alpha=1$, and by expanding it about $1/\bar{\lambda}_2^{1/5}$. 

Let us discuss the insensitivity of the relation to variation in $\alpha$. Substituting the expansion of ${\cal C}$ in $\bar{\lambda}_2$ into Eq.~\eqref{eq:barlambdainC} with $\ell=3$ without fixing~$\alpha$, one finds
\begin{align}
    \bar{\lambda}_3= \bar{\lambda}_2^{7/5} \left[C_0^{\bar{\lambda}_3 \bar{\lambda}_2}+\frac{C_1^{\bar{\lambda}_3 \bar{\lambda}_2}}{\bar{\lambda}_2^{1/5}} +{\cal O}\left(\frac{1}{\bar{\lambda}_2^{2/5}}\right) \right],\label{eq:AnalyticMultipoleLove}
\end{align}
where $C_0^{\bar{\lambda}_3 \bar{\lambda}_2}=C_0^{\bar{\lambda}_3 \bar{\lambda}_2}\left(\alpha\right)$ and $C_1^{\bar{\lambda}_3 \bar{\lambda}_2}=C_1^{\bar{\lambda}_3 \bar{\lambda}_2}\left(\alpha\right)$. The Taylor expansion of $C_0^{\bar{\lambda}_3 \bar{\lambda}_2}$ and $C_1^{\bar{\lambda}_3 \bar{\lambda}_2}$ around $\alpha=1$ reads
\begin{align}
    C_0^{\bar{\lambda}_3 \bar{\lambda}_2}=& 0.16\left[1+0.066 \left(\alpha-1\right)+{\cal O}\left(\left(\alpha-1\right)^2\right)\right],\nonumber\\
     C_1^{\bar{\lambda}_3 \bar{\lambda}_2}=& 6.6\left[1+ 0.19 \left(\alpha-1\right)+{\cal O}\left(\left(\alpha-1\right)^2\right)\right]. 
\end{align}
For the leading-order contribution, the first-order coefficient in the expansion is less than $10\%$ of the zeroth-order coefficient, indicating the weak~$\alpha$ dependence. We assess the universality by
\begin{align}
    \frac{\left. C_0^{\bar{\lambda}_3 \bar{\lambda}_2} \right|_{\alpha=0}}{\left. C_0^{\bar{\lambda}_3 \bar{\lambda}_2}\right|_{\alpha = 1.4}}\simeq 0.92.\label{eq:RatioinEven2Even3}
\end{align}
This means the deviation of the $\bar{\lambda}_2-\bar{\lambda}_3$ relation in the range of $\alpha$ under consideration is $8\%$, which is, once again, much smaller than the deviation between polytropes with $0\le n \le1$, found as $20\%$~\cite{Yagi:2013sva}.

\section{Potential origin of the approximate universality}\label{Section:PotentialOrigin}
The power-law relations are characterized by the following two key aspects: the independence of the power-law exponent from the details of microscopic physics and the weak dependence of the power-law coefficients on these details. In the following, we discuss these one by one, by focusing on the leading-order contribution of the PM expansion, which sufficiently captures the approximate universal behavior. After that, we argue that the stellar structures with low compressibility appear to play a crucial role in universality.

\subsection{EoS independence of the power-law exponent }
Equations~\eqref{eq:barlambdainC} and~\eqref{eq:barsigmainC} imply that $\bar{\lambda}_\ell$ and $\bar{\sigma}_\ell$ scale as ${\cal C}^{-2\ell-1}$ and ${\cal C}^{-2\ell}$, respectively, to leading order of the PM expansion. A comparison by eliminating ${\cal C}$ from them reveals the scaling relation, $\bar{\sigma}_\ell \propto \bar{\lambda}_\ell^{2\ell/(2\ell+1)}$, which determines the exponent in Eq.~\eqref{eq:AnalyticEvenOddLove} for $\ell=2$. Similarly, one finds $\bar{\lambda}_{\ell'} \propto \bar{\lambda}_\ell^{(2\ell'+1)/(2\ell+1)}$ for $\ell'(\neq \ell)$, explaining the exponent in Eq.~\eqref{eq:AnalyticMultipoleLove} for $(\ell,\ell')=(2,3)$. The exponent is thus determined by the scaling law of the tidal deformabilities with respect to ${\cal C}$, which is independent of the underlying microphysics.

\subsection{Suppression of EoS dependence of the power-law coefficients}
We here mainly discuss the Even-Odd Love relation~\eqref{eq:lambda2sigma2}, as it shares fundamentally same structures with the multipole Love relation~\eqref{eq:lambda2lambda3}. Eliminating ${\cal C}$ from Eqs.~\eqref{eq:electriclambda} and~\eqref{eq:magneticsigma} (noting $j_\ell={\cal O}({\cal C}^1$)) at leading order of the PM expansion with $\ell=2$, we obtain
\begin{align}
  \bar{\sigma}_2 \propto \frac{j_2^{\rm (1PM)}}{(k_2^{\rm (N)})^{4/5}} \bar{\lambda}_2^{4/5},\label{eq:EvenOddatleading}
\end{align}
where $k_2^{\rm (N)}$ and $j_2^{\rm (1PM)}$ are the Newtonian quadrupolar TLN and 1PM-order magnetic-type quadrupolar TLN, respectively. The factor in front of $\bar{\lambda}_2^{4/5}$ on the right-hand side shares the~$\alpha$ dependence with $C_0^{\bar{\sigma}_2 \bar{\lambda}_2}$ in Eq.~\eqref{eq:AnalyticEvenOddLove}; the only difference is an irrelevant numerical factor. 

Let us first discuss the $\alpha$ dependence of $k_2^{\rm (N)}$ and $j_2^{(\rm 1PM)}$. We expand them around $\alpha=\alpha_0 (\in [0,1.4])$, obtaining
\begin{align}
 k_2^{(\rm N)}= & C_{k_2}^{(0)}\left[1-C_{k_2}^{(1)} \left(\alpha-\alpha_0\right)+{\cal O}\left(\left(\alpha-\alpha_0\right)^2\right)\right],\label{eq:k2expansionaroundalpha0}\\
 j_2^{(\rm 1PM)}=& -  C_{j_2}^{(0)}  \left[1-C_{j_2}^{(1)} \left(\alpha-\alpha_0\right)+{\cal O}\left(\left(\alpha-\alpha_0\right)^2\right)\right],\label{eq:j2expansionaroundalpha0}
\end{align}
where $C_{k_2}^{(i)}$ and  $C_{j_2}^{(i)}$ are positive numbers. It is worth mentioning that the relative change rates of $k_2^{\rm (N)}$ and $j_2^{\rm (1PM)}$ with respect to $\alpha$, i.e., $-C_{k_2}^{(1)}$ and $-C_{j_2}^{(1)}$, are negative. Physically, this implies that smaller~$\alpha$~(stiffer EoS) leads to stronger tidal deformation. The magnitude of $C_{k_2}^{(1)}$ and  $C_{j_2}^{(1)}$ quantifies the $\alpha$ dependence of $ k_2^{(\rm N)}$ and $ j_2^{(\rm 1PM)}$, respectively. For example, $(C_{k_2}^{(1)},C_{j_2}^{(1)})\simeq (0.31,0.21)$ for $\alpha_0=1$. We found $0.2\lesssim C_{k_2}^{(1)} \lesssim 0.5$ and $C_{j_2}^{(1)}\simeq 0.68 C_{k_2}^{(1)}$ (the latter to be explained below) for $0\le \alpha_0 \le 1.4$, implying that the $\alpha$ dependence of $k_2^{\rm (N)}$ and $j_2^{\rm (1PM)}$ is not significant but slightly stronger than that in the universal Love relations derived in Section~\ref{Sec:ApproximateUbniversalRelations}. On the other hand, the~$\alpha$ dependence of the ratio,\footnote{We reinterpret $\alpha_0$ as $\alpha$ when discussing the dependence of relevant quantities on $\alpha.$ }
\begin{equation}
\label{eq:ratio}
    r_{\bar \lambda_2 - \bar \sigma_2}^{(1)} (\alpha) \equiv \frac{C_{j_2}^{(1)} (\alpha)}{C_{k_2}^{(1)} (\alpha)},
\end{equation}
in $0\le \alpha \le 1.4$ is evaluated as 
\begin{align}
\frac{r_{\bar \lambda_2 - \bar \sigma_2}^{(1)}(0)}{r_{\bar \lambda_2 - \bar \sigma_2}^{(1)}(1.4)} \simeq 0.97, \label{eq:rarioofC1}
\end{align}
which shows the insensitivity to $\alpha$. The ratio, $C_{j_2}^{(1)}/C_{k_2}^{(1)} \simeq 0.68$, thus approximately holds in $0\le \alpha \le 1.4$. We have also confirmed that the ratio deviates from $0.68$ as $\alpha$ further increases. Physically, Eq.~\eqref{eq:rarioofC1} implies the similarity of the $\alpha$ dependence of $k_2^{\rm (N)}$ and $j_2^{\rm (1PM)}$ in $0\le \alpha \le 1.4$.
Another intriguing finding is that the ratio is close to the value of the power-law exponent,~$4/5~(=0.8)$, which appears to play a certain role in suppressing the $\alpha$ dependence as will be seen below. 

 The similar structure can be found in other universal Love relations as well, which is summarized in Table~\ref{Table:ratio}, and even in the I-Love relation as discussed in Appendix~\ref{Appendix:ILove}. The physical origin of these structures remains unclear; we discuss it again at the end of this section.
\begin{table*}[t]
\begin{tabular}{ |c|c|c|c| } 
\hline
$x-y$ relation &  Ratio $r_{x-y}^{(1)}(1)$ & Exponent $\bar n$ & Ratio $r_{x-y}^{(1)}(0)/r_{x-y}^{(1)}(1.4)$\\
\hline
$\bar{\lambda}_2- \bar{\sigma}_2$ & 0.68 & 4/5~(=0.8) & 0.97  \\ 
\hline
$\bar{\lambda}_2- \bar{\lambda}_3$ & 1.1  & 7/5~(=1.4) & 1.1\\ 
\hline
$\bar{\sigma}_2- \bar{\sigma}_3$ & 1.2 & 3/2~(=1.5) & 0.94\\ 
\hline
\end{tabular}
\caption{The ratio of the relative change rates of TLNs, defined in the same manner as Eq.~\eqref{eq:ratio},
the power-law exponent $\bar n$ of each universal relation defined as $y \propto x^{\bar n}$ to leading order in the PM expansion, and the assessment of the insensitivity of the ratio to $\alpha$, similar to Eq.~\eqref{eq:rarioofC1}. Observe that the ratio $r_{x-y}^{(1)}(1)$ takes a value similar to the power-law exponent. Note that the ratio deviates from those values as $\alpha$ further increases. }
\label{Table:ratio}
\end{table*}

\begin{figure}[htbp]
\centering
\includegraphics[scale=0.64]{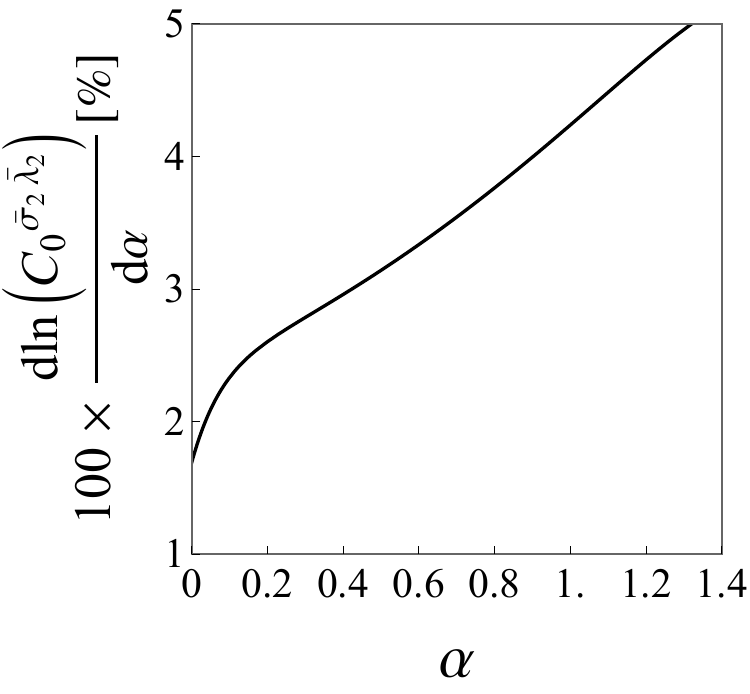}
\hspace{0.3cm}
\caption{The relative change rate of $C_0^{\bar{\sigma}_2 \bar{\lambda}_2}$ in Eq.~\eqref{eq:AnalyticEvenOddLove} with respect to $\alpha$, which corresponds to the ratio of the first-order coefficient to the zeroth-order one in the expansion~\eqref{eq:powerlawcoefficientatleadingPM}. Observe that the rate is less than $5\%$ in $0\le \alpha \lesssim  1.4$, indicating approximate universality between $\bar{\lambda}_2$ and $\bar{\sigma}_2$.}
\label{Fig:dCsigmalambda}
\end{figure}

Let us now discuss the combination of TLNs. Equation~\eqref{eq:EvenOddatleading} shows that $k_2^{\rm (N)}$ contributes to the power-law coefficient with an inverse power. This means that the $\alpha$-dependent contribution arising from $k_2^{\rm (N)}$ has the opposite sign to that from $j_2^{\rm (1PM)}$ in the combination, and thus, the total $\alpha$ dependence in the coefficient is diminishing with each other. Indeed, one can derive
\begin{align}
     \frac{j_2^{\rm (1PM)}}{(k_2^{\rm (N)})^{4/5}} \propto  1-\left(C_{j_2}^{(1)}-\frac{4}{5}C_{k_2}^{(1)}\right)\left(\alpha-\alpha_0\right)  +{\cal O}\left(\left(\alpha-\alpha_0\right)^2 \right).\label{eq:powerlawcoefficientatleadingPM}
\end{align}
Since $C_{j_2}^{(1)}\simeq 0.68 C_{k_2}^{(1)}$ and $C_{k_2}^{(1)}={\cal O}(10^{-1})$ as mentioned previously, the factor in front of $(\alpha-\alpha_0)$ on the right-hand side is of ${\cal O}(10^{-2})$, implying that the $\alpha$ variation in the power-law coefficient in Eq.~\eqref{eq:EvenOddatleading} is of order of $1\%$, as demonstrated in Fig.~\ref{Fig:dCsigmalambda}. Thus, the particular combination of TLNs suppresses their EoS dependence, leading to approximate universality. Interestingly, Fig.~\ref{Fig:dCsigmalambda} shows that the $\alpha$ dependence tends to weaken as $\alpha$ decreases, reaching a minimum at $\alpha=0$. This implies that~(i)~$C_0^{\bar{\sigma}_2 \bar{\lambda}_2}$ in Eq.~\eqref{eq:AnalyticEvenOddLove} is a monotonically increasing function of $\alpha$, explaining why the ratio in Eq.~\eqref{eq:RatioinEvenOdd} is smaller than unity;~(ii)~the universality is continuously reinforced as the local adiabatic index in the entire interior increases. The multipole Love relation~\eqref{eq:AnalyticMultipoleLove} can also be understood in the same way. 

An intriguing observation in the suppression mechanism is that the ratio of the relative change rates of TLNs with respect to $\alpha$ is a similar value to the power-law exponent in the power-law relations, as presented in Table~\ref{Table:ratio}. In Appendix~\ref{Appendix:ILove}, we find the same structure in the I-Love relation as well~(see Table~\ref{Table:ratioforILove} therein). This structure plays a certain role in universality, as the ratio deviates from those values as $\alpha$ further increases, which is consistent with Fig.~\ref{Fig:dCsigmalambda}. Again, the insensitivity of the rate to $\alpha$ suggests the similarity of the $\alpha$ dependence of TLNs. The origin of the number of the ratio still remains unclear; we leave further considerations about the physical origin for future work.

In summary, the approximate universal Love relations rely on the suppression of the EoS dependence in a particular combination of TLNs. This suppression mechanism essentially rests on the macroscopic property of tidally deformed NSs: the absolute values of TLNs increase for stiffer EoSs, implying that stiffer EoSs lead to stronger tidal deformation. Note that the modest EoS dependence of TLNs is not generically a necessary condition for the universal Love relations despite being closely related to their assessment, as argued in the next section.

\subsection{Modest EoS dependence of TLNs}\label{Section:LoveC}
\begin{figure}[htbp]
\centering
\includegraphics[scale=1.0]{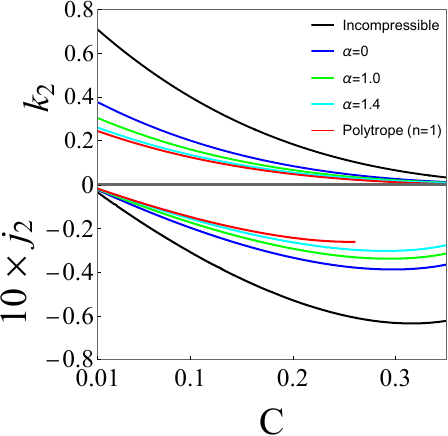}
\hspace{0.3cm}
\caption{The $k_2$-$j_2$-${\cal C}$ relation. For a given EoS, a set of TLNs form a curve as a function of ${\cal C}$. The incompressible limit~(black curves) corresponds to the outer edge of the region covered by a set of EoS curves in the parameter space that keeps pressure finite. Observe that, for larger~$\alpha$~(softer EoS), the region surrounded by the curves shrinks.}
\label{Fig:Evenk2Oddj2}
\end{figure}

One might expect that the suppression mechanism previously discussed relies on the modest EoS dependence of $k_\ell$ and $j_\ell$. We now argue that this is not a necessary condition for the universal Love relations despite being closely related to their assessment. Figure~\ref{Fig:Evenk2Oddj2} implies that the variation in $k_\ell$ with respect to $\alpha$ in $0\le \alpha \le 1.4$ is small, compared to polytropes with $0\le n \le 1$~(see, e.g., Fig.~2 in Ref.~\cite{Postnikov:2010yn}). Indeed, the EoS dependence is assessed by
\begin{align}
    \frac{k_2^{\rm (N)}|_{\alpha=0}}{k_2^{\rm (N)}|_{\alpha=1.4}}\simeq 1.4,\quad
    \frac{j_2^{\rm (1PM)}|_{\alpha=0}}{j_2^{\rm (1PM)}|_{\alpha=1.4}}\simeq 1.3,\label{eq:RatioofTLNs}
\end{align}
and
\begin{align}
    \frac{k_2^{\rm (N)}|_{n=0}}{k_2^{\rm (N)}|_{n=1}}\simeq 2.9,\quad
    \frac{j_2^{\rm (1PM)}|_{n=0}}{j_2^{\rm (1PM)}|_{n=1}}\simeq 2.1.
\end{align}
The former deviation is of ${\cal O}(10)\%$, while the latter is of ${\cal O}(10^2)\%$. On the one hand, the modest~$\alpha$ dependence in the modified Tolman~VII solution may be related to the Love-C relation, discovered in NSs with realistic EoSs~\cite{Maselli:2013mva,Yagi:2016bkt} as well as the fully analytic model of the modified Tolman~VII solution~\cite{Jiang:2020uvb}. On the other hand, tidal deformabilities (and TLNs) for polytropes with $0\le n \le1$ sensitively depend on $n$ but follow the universal Love relations due to the suppression mechanism in a weaker manner than both the modified Tolman~VII solution and realistic EoSs~\cite{Yagi:2013sva}.

The universal Love relations hold more robustly than the~$k_\ell-{\cal C}$ and $j_\ell-{\cal C}$ relations due to the suppression mechanism that works among macroscopic physical parameters induced by perturbations. It is important to note that the~$k_\ell-{\cal C}$ and $j_\ell-{\cal C}$ relations do not underlie the universal Love relations, nor vice versa; rather, they should be regraded as mutually correlated in a non-hierarchical manner. The~$k_\ell-{\cal C}$ and $j_\ell-{\cal C}$ relations appear to originate from an approximate quadratic energy density profile shared by NSs with realistic EoSs and the modified Tolman~VII solution~\cite{Jiang:2020uvb,Yagi:2013bca}. This is consistent with the fact that TLNs for polytropes with $0\le n \le 1$ strongly depend on $n$. However, the universal Love relations still hold even for the polytropes, suggesting that the suppression mechanism arises from different structures from the similarity of the energy density profile. This is more evident in the I-Love relation, as seen in Appendix~\ref{Appendix:ILove}.

\subsection{Universality from low compressibility}\label{Section:UniversalityfromStellarStructure}
Figure~\ref{Fig:dCsigmalambda} suggests that smaller~$\alpha$~(stiffer EoS) reinforces universality. This seems to be consistent with the fact that universality becomes less tight as one continuously increases the polytropic index in the regime $n > 1$.
Therefore, it is natural to expect that universality may arise from stellar structures associated with low compressibility. 

\begin{figure}[htbp]
\centering
\includegraphics[scale=0.60]{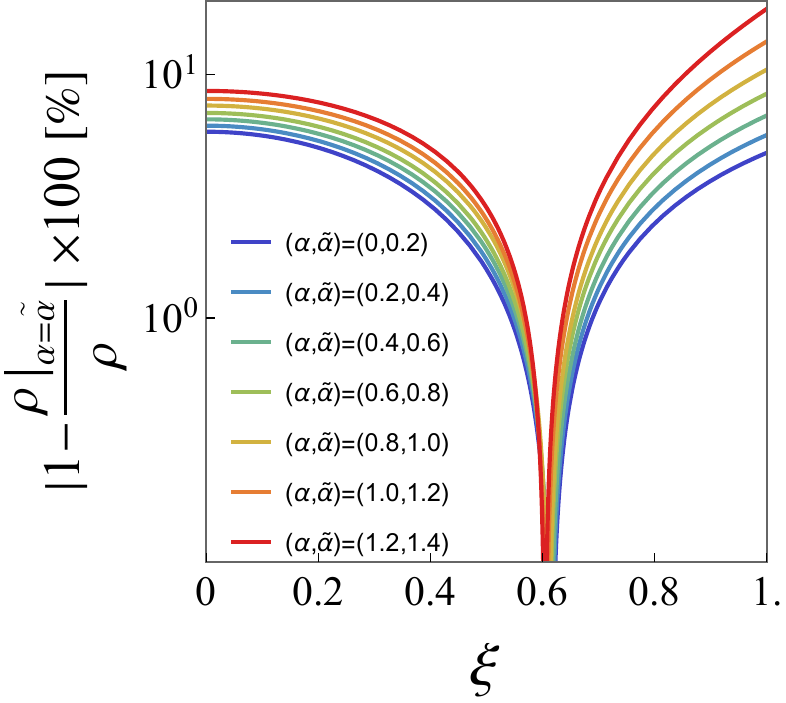}
\hspace{0.3cm}
\caption{The fractional difference in $\rho$ given by Eq.~\eqref{eq:densityofmTolman} for similar values of $\alpha$ with $\bar{\lambda}_2=1000$, as a function of $\xi(=r/R)$. Observe that, for larger~$\alpha$, the fractional difference increases in a large portion of the interior.}
\label{Fig:FracDiff_EnergyDensity}
\end{figure}

We find that the trend in the $\alpha$ dependence of the power-law coefficients is inherited from that of the equilibrium stellar configuration, i.e., the modified Tolman~VII solution. Figure~\ref{Fig:FracDiff_EnergyDensity} presents the fractional difference in the energy density profiles~\eqref{eq:densityofmTolman} for similar values of $\alpha$ with a fixed $\bar{\lambda}_2=1000$, showing that the difference is typically less than $10\%$ in a large potion of the interior and increases for larger~$\alpha$. Similar behavior can be found in other background quantities, such as $p$, $\nu$, and~$m$. In other words, the stellar configuration in equilibrium tends to become insensitive to $\alpha$ variations as $\alpha$ approaches zero. The ``valley'' observed in Fig.~\ref{Fig:FracDiff_EnergyDensity} originates from the crossing of the energy-density profiles for different values of $\alpha$ around $\xi=0.6$, as shown in Fig.~\ref{Fig:EnergyDensityProfile}.

Building upon the above considerations, we expect that low compressibility of stellar configurations in the entire interior region plays a crucial role in approximate universality for NSs. These observations suggest the following picture: incompressible stars correspond to a restricted subset of equilibrium stellar configurations in the phase space spanned by physical degrees of freedom, as the energy density remains constant throughout the entire interior. Linear perturbations to such incompressible stars induce the strongest possible responses, setting upper bounds on their magnitudes. Compressible stars deviate from this limit by decreasing the adiabatic index. For NSs, stellar configurations remain insensitive to variations in the EoS due to low compressibility. A decrease in the adiabatic index reduces the magnitudes of the responses in a manner commonly shared among various tidal perturbations, leading to approximate universality. Eventually, as the adiabatic index decreases further, the degrees of freedom associated with an EoS expand, and in turn, the similarity of the responses is lost, resulting in the loss of universality.

\subsection{Comparison with previous findings}

Let us discuss consistencies with findings from previous studies. Reference~\cite{Sham:2014kea} demonstrated that the I-Love relation is most strongly reinforced around the incompressible limit~(polytrope with~$n=0$) using an exact solution to Eq.~\eqref{eq:EinsteinEqs}, called a generalized Tolman~VII model. It is worth noting that this model serves as a good approximation for quark stars rather than NSs, except in the Tolman~VII-solution limit as the density is generally discontinuous at the stellar surface. Within this framework, the incompressible limit corresponds to a stationary point of the I-Love relation, in the sense that the relative change rate of the coefficient in the power-law relation becomes exactly zero~(see Fig.~5 in the reference). In Appendix~\ref{Appendix:ILove}, we demonstrate that the rate for the I-Love relation within our framework reaches a minimum in the locally incompressible limit in a similar manner to the universal Love relations. 

The proposal by Ref.~\cite{Sham:2014kea} is that the I-Love-Q relation can be attributed to the incompressible limit; the high stiffness of EoSs causes approximate universality of NSs. Establishing this statement requires addressing a question regarding the difference in the profile of the local adiabatic index in the generalized Tolman~VII model from that of realistic EoSs for NSs, particularly in the outer region. In the previous section, we found the strongest reinforcement of the universal Love relations~(as well as the I-Love relation, as discussed in Appendix~\ref{Appendix:ILove}) around the locally incompressible limit in the inner core, which is consistent with this statement. Therefore, our findings support and complement the claim in Ref.~\cite{Sham:2014kea} by providing a comprehensive perspective on the approximate universality of NSs, based on an analytic model for realistic NSs, instead of quark stars. 

Reference~\cite{Yagi:2014qua} found that the isodensity contours of rotating NSs with realistic EoSs can be approximated by self-similar ellipsoids and relaxing this assumption in stellar modeling destroys the three hair relations discovered by Ref.~\cite{Stein:2013ofa}. These results suggest that the self-similarity of elliptical stellar isodensity surfaces is responsible for the approximate universality. Figure~17 in Ref.~\cite{Yagi:2014qua} demonstrates that the spatial variation in stellar eccentricity of Newtonian polytropes tends to weaken as the polytropic index~$n$ decreases, and vanishes in the incompressible fluid case~($n=0$). This suggests that, for stiffer EoSs, the energy density profile is better approximated by a sequence of self-similar surfaces with constant stellar eccentricity. Notably, this approximation becomes exact when $n=0$. On the other hand, softer EoSs tend to break this self-similar condition, resulting in the loss of universality. These findings further highlight the importance of stiffness of EoSs in approximate universality, which is consistent with our results.

\section{Discussion and extensions}\label{Section:Discussion}
We have conducted a comprehensive theoretical investigation of approximate universality of NSs using a semi-analytic approach. Our analysis relies on the semi-analytic NS interior model proposed in Ref.~\cite{Posada:2022lij} and fully analytic NS interior model proposed in Ref.~\cite{Jiang:2019vmf}, both of which approximately recover static equilibrium configurations of NSs with realistic EoSs in a wide range of compactnesses. The derived power-law relations among tidal deformabilities agree well with the empirical fit in Ref.~\cite{Yagi:2013sva}. 

In general, magnetic-type tidal deformabilities are expected to be difficult to measure, even in third generation GW observations due to their small magnitude compared to the even-parity tidal deformabilities. However, their sign depends on the fluid state in the NS interior~\cite{Landry:2015cva,Pani:2018inf}, which may leave detectable imprints in the GW phase~\cite{JimenezForteza:2018rwr}. Our analysis of the odd-parity stellar perturbations assumes an irrotational state that permits internal motions, leading to negative-valued magnetic-type tidal deformabilities, while assuming a strictly static state results in positive values~\cite{Binnington:2009bb}. The universal Love relations derived in this work can be used to infer the fluid state inside realistic NSs within GR.

A potential extension of the present work involves incorporating the effect of NS rotation, even though NSs in a binary system are generically expected to rotate slowly. The perturbative framework in Refs.~\cite{Pani:2015hfa,Landry:2015zfa} about a small spin parameter  shows that the coupling between NS spin and a tidal environment gives rise to a new class of tidal coefficients, dubbed rotational TLNs. Notably, it has been suggested that slowly spinning NSs do not exhibit the I-Love relations with respect to the rotational TLNs~\cite{Pani:2015nua}. Exploring the breakdown of universality due to the spin-tidal coupling would provide a deeper insight into the fundamental physical origin of universality. Alternatively, it is beneficial to study whether universality with respect to rotational TLNs is preserved under a different construction of relevant dimensionless quantities, as the strength of universality appears to depend on normalization of physical parameters~\cite{Majumder:2015kfa}. 

From a practical perspective, if the universal Love relations hold even for rotational TLNs, they could provide a powerful tool for waveform modeling. This is because the leading-order rotational TLN enters at $6.5$PN order, which is lower than the $\ell\ge 3$rd electric-type and magnetic-type tidal deformabilities~\cite{JimenezForteza:2018rwr}. Furthermore, neglecting the spin-tidal coupling may introduce a significant bias in parameter estimation for NSs with dimensionless spin of $\chi\sim 0.1$ with  third-generation detectors~\cite{JimenezForteza:2018rwr,Castro:2022mpw}. Establishing an EoS insensitive connection between the rotational TLN and the quadrupolar electric-type tidal deformability could enhance the ability to infer the EoS at supernuclear densities and test GR in the strong-field regime.

This work focuses on NSs within GR. Modifying the theory of gravity can alter the description of the coupling between gravity and matter, leading to deviations in tidal deformabilities from their GR values. It is not obvious whether NSs in modified theories of gravity exhibit approximate universality among various tidal deformabilities. If universal Love relations hold even for non-GR theories, the power-law relations would differ from those in GR. The EoS insensitive property provides a novel approach to test strong gravity through tidal effects in GW observations.

The approximate universality of tidally deformed NSs is reminiscent of a hidden symmetry underlying the vanishing TLNs of black holes~\cite{Hui:2021vcv,Katagiri:2022vyz,Charalambous:2022rre} in the sense that Schwarzschild black holes have identical~(zero) TLNs at any multipole orders in both the even-parity and odd-parity sectors. Since the exterior region of a nonspinning NS is uniquely described by the Schwarzschild spacetime, one could interpret that the compact stellar interior region weakly breaks the symmetry, leading to nonzero TLNs, whose values increase as the stellar compactness decreases. Although there is no stationary continuous sequence from a NS to a black hole within a perfect-fluid approximation due to the Buchdahl limit~\cite{Buchdahl:1959zz}, the approximate universal Love relations could be understood in this context. We leave further consideration for future study.

\acknowledgments
We are indebted to Zexin Hu and Lijing Shao for their valuable comments and insights. T.K. and G.R.M. acknowledge support from the Villum Investigator program supported by the VILLUM Foundation (grant no. VIL37766) and the DNRF Chair program (grant no. DNRF162) by the Danish National Research Foundation. This work has received funding from the European Union’s Horizon 2020 research and innovation programme under the Marie Sklodowska-Curie grant agreement No 101131233. The Center of Gravity is a Center of Excellence funded by the Danish National Research Foundation under grant No. 184.
K.Y. acknowledges support from NSF Grant No. PHY-2309066, No. PHYS-2339969, and the Owens Family Foundation.

\appendix

\section{Modified Tolman~VII model}\label{Appendix:modifiedTolmanVII}

\subsection{Semi-analytic model}
\begin{figure}[htbp]
\centering
\includegraphics[scale=0.53]{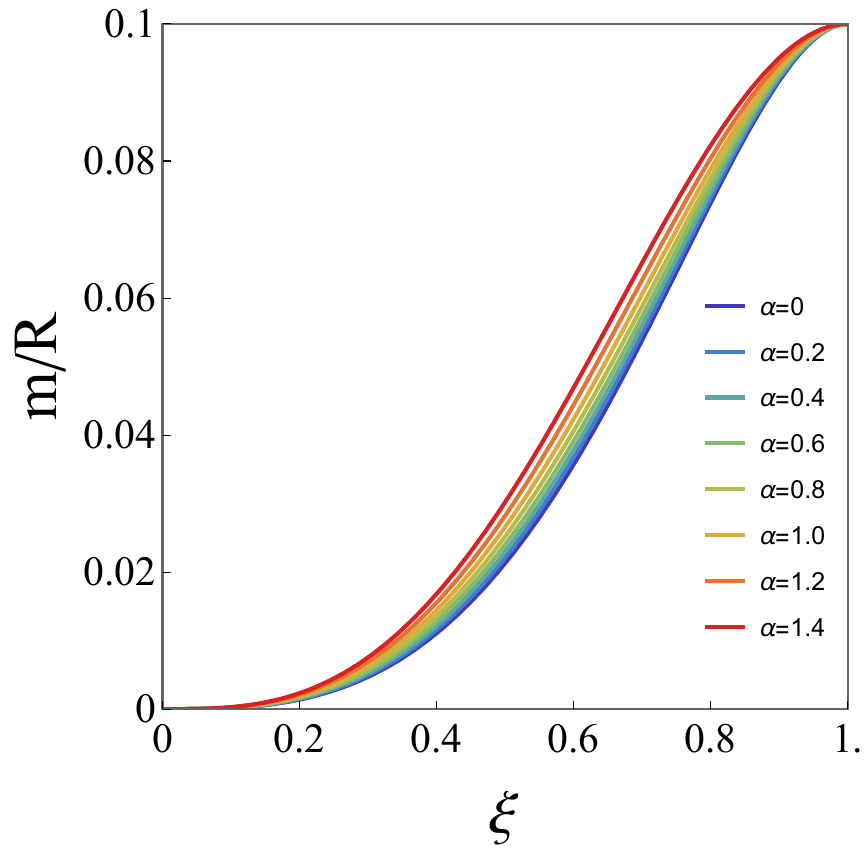}
\hspace{0.3cm}
\caption{The mass function profile of the modified Tolman~VII solution.
}
\label{Fig:massmTolman}
\end{figure}

\begin{figure}[htbp]
\centering
\includegraphics[scale=0.55]{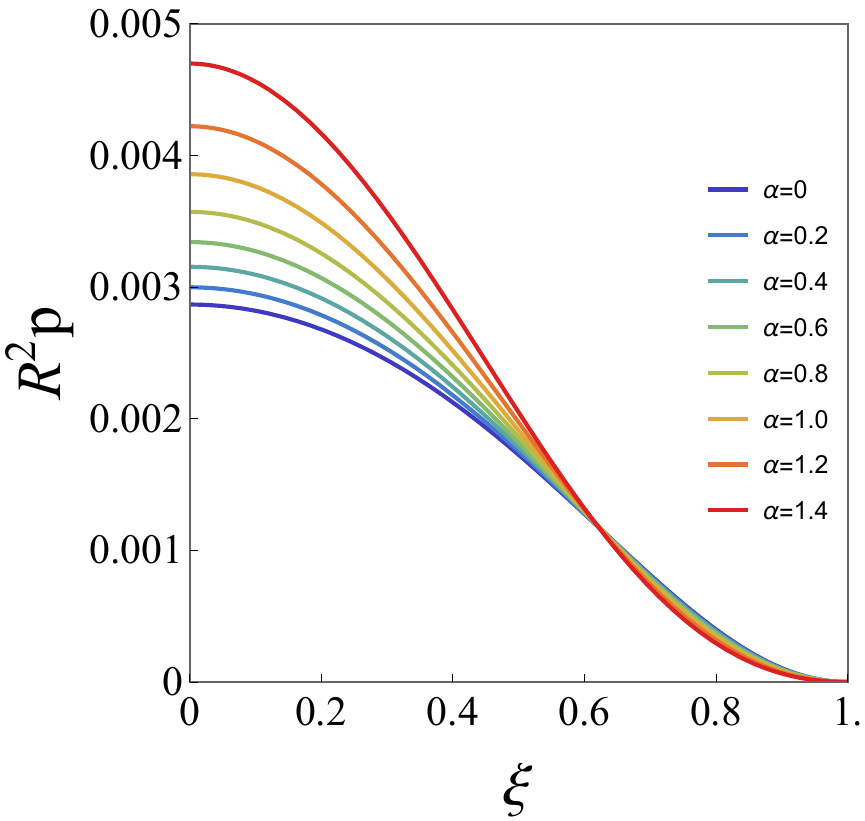}
\hspace{0.3cm}
\caption{The pressure profile of the modified Tolman~VII solution.
}
\label{Fig:pmTolman}
\end{figure}

\begin{figure}[htbp]
\centering
\includegraphics[scale=0.54]{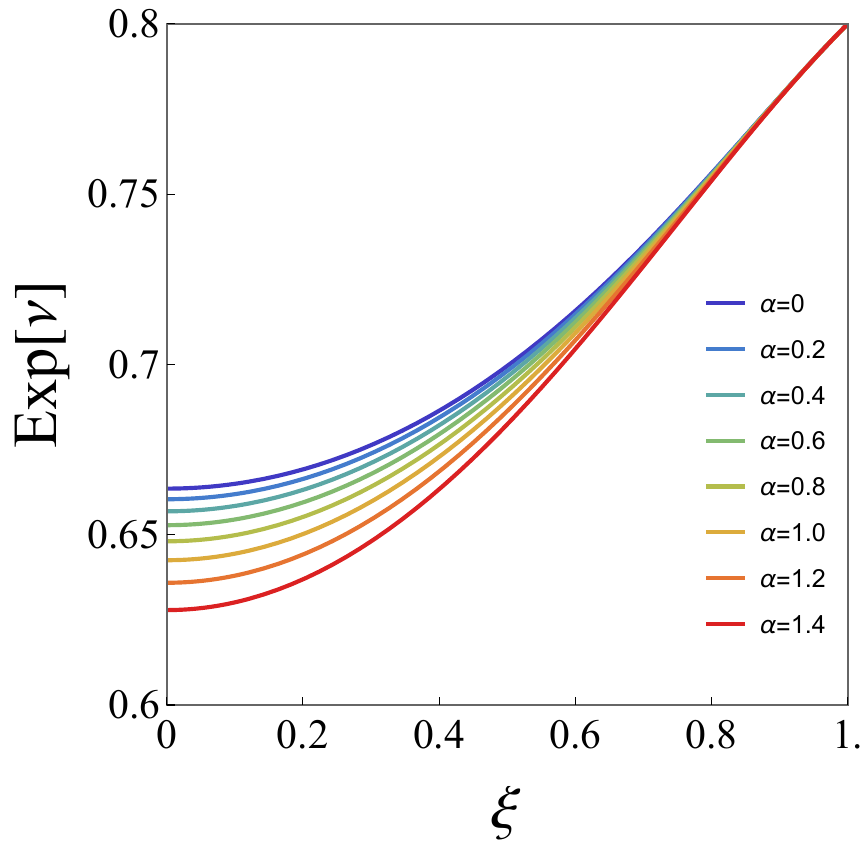}
\hspace{0.3cm}
\caption{The lapse function profile of the modified Tolman~VII solution.
}
\label{Fig:lapsemTolman}
\end{figure}

We provide the modified Tolman~VII solution in Figs.~\ref{Fig:massmTolman} --~\ref{Fig:lapsemTolman} with ${\cal C}=0.1$. The energy density profile is given by Eq.~\eqref{eq:densityofmTolman} and is depicted in Fig~.\ref{Fig:EnergyDensityProfile}.

\subsection{Fully analytic model}
The explicit form of the fully analytic model of the modified Tolman~VII solution is given by~\cite{Jiang:2019vmf}
\begin{align}
    e^\nu=&C_1 \cos^2\phi,  \\
    e^{-\lambda}=& 1-8\pi \rho_c R^2 \xi^2 \left(\frac{1}{3}-\frac{\alpha}{5}\xi^2+\frac{\alpha-1}{7}\xi^4\right),\nonumber
\end{align}
with
\begin{align}
    \phi=&C_2 -\frac{1}{2}\ln \left(\xi^2-\frac{5}{6}+\sqrt{\frac{5 e^{-\lambda|_{\alpha=1}}}{8\pi \rho_c R^2}}\right),\nonumber\\
    C_1=& \left(1-\frac{2M}{R}\right)  \nonumber \\
    &\times\left\{1+\frac{8\pi \rho_c R^2\left(10-3\alpha\right)^2\left(15-16\pi \rho_c R^2\right)}{3\left[105+16\pi \rho_c R^2\left(3\alpha-10\right)\right]^2}\right\} ,\\
    C_2=&\arctan\left[-\frac{2\left(10-3\alpha\right)R\sqrt{6\pi \rho_c \left(15-16\pi \rho_c R^2\right)}}{48\pi \rho_c \left(10-3\alpha\right)R^2-315}\right]\nonumber\\
    &+\frac{1}{2}\ln \left(\frac{1}{6}+\sqrt{\frac{5}{8\pi \rho_c R^2}-\frac{2}{3}}\right), \nonumber
\end{align}
and 
\begin{align}
        p=& \frac{\tan\phi}{R}\sqrt{\frac{\rho_c}{10\pi} \left[1-\frac{8\pi\rho_c R^2}{15}\xi^2\left(5-3\xi^2\right) \right]}\label{eq:rhoandponBG}\\
    &-\frac{\rho_c}{15}\left(5-3\xi^2\right)+\frac{6\left(1-\alpha\right)\rho_c}{16\pi \rho_c R^2\left(10-3\alpha\right)-105}.\nonumber
\end{align}
Here, $\rho_c$ is the energy density at the stellar origin, related to the total mass~$M$ via 
\begin{align}
    \rho_c=\frac{105M}{8\pi R^3 \left(10-3\alpha\right)}.\label{eq:rhoc}
\end{align}
This model satisfies boundary conditions~$e^{\nu(R)}=e^{-\lambda(R)}=1-2M/R$ and $p(R)=\rho(R)=0$ with $m(R)=M$.

\section{Linear perturbations for relativistic stars}\label{Appendix:LinearStellarPerturbation}

In this appendix, we describe linear perturbations for relativistic stars. We explain even-parity and odd-parity sectors separately.

\subsection{Even-parity sector}

\subsubsection{Stellar perturbations}
We first briefly introduce the general framework of frequency-dependent linear perturbations of relativistic stars. In the Regge-Wheeler gauge~\cite{PhysRev.108.1063}, the even-parity sector of metric perturbations with harmonic decomposition takes the form,
\begin{align}
       &h_{\mu \nu}^{+}dx^\mu dx^\nu =  ( e^\nu H^{\ell m} dt^2+2H_1^{\ell m}dt dr\label{eq:heven}\\
    &+e^\lambda H_2^{\ell m}dr^2+r^2 K^{\ell m}d\Omega^2)e^{-i \omega t} Y_{\ell m}.\nonumber
\end{align}
Here, $H^{\ell m}=H^{\ell m}(r)$, $H_1^{\ell m}=H_1^{\ell m}(r)$, $H_2^{\ell m}=H_2^{\ell m}(r)$, $K^{\ell m}=K^{\ell m}(r)$; $Y_{\ell m}=Y_{\ell m}(\theta, \varphi)$ are spherical harmonics. Henceforth, we drop the superscript~$\ell m$.

The even-parity metric perturbation induces the displacement of $\rho$, $p$, and the four velocity. We parametrize the Lagrangian displacement of the fluid variables by two functions, $W=W(r)$ and $V=V(r)$, so that the components of the displacement vector~$\xi^\mu$ are expressed by~\cite{1983ApJS...53...73L}
\begin{align}
    \xi^r=&r^{\ell-1} e^{-\lambda/2} W e^{-i \omega t} Y_{\ell m},\quad
    \xi_A=-r^\ell V e^{-i \omega t} E_A^{\ell m},
\end{align}
where $E_A^{\ell m}:=(0,0,\partial_\theta Y_{\ell m},\partial_\varphi Y_{\ell m})$ are even-parity vector spherical harmonics. We assume $u_\mu \xi^\mu=0$ by using the gauge degrees of freedom. One can express the Eulerian perturbation of the four velocity, $\delta u^\mu$, in terms of $W$ and $V$ as follows. First, for the Lagrangian variation~$\Delta$, using the normalization condition $(g_{\mu\nu}+\Delta g_{\mu\nu})(u^\mu+\Delta u^\mu)(u^\nu+\Delta u^\nu)\simeq -1$ to linear order, and noticing that $\Delta u^\mu$ should be proportional to $u^\mu$ linearly, one obtains the expression for $\Delta u^\mu$ by~\cite{Friedman:1978wla}
\begin{align}
    \Delta u^\mu =\frac{1}{2} u^\mu u^\alpha u^\beta \Delta g_{\alpha \beta}.\label{eq:Deltaumu}
\end{align}
Next, using the relation that relates the Lagrangian variation~$\Delta$ with the Eulerian variation $\delta$ through $\Delta=\delta+{\cal L}_\xi$, where ${\cal L}_\xi$ is the Lie derivative along $\xi$, one obtains
\begin{align}
    &\delta  u^t=\frac{1}{2} e^{-\nu/2} H e^{-i \omega t} Y_{\ell m},\\
    &\delta u^r= - i \omega e^{-\nu/2} \xi^r,\quad   \delta u^A=  -i \omega e^{-\nu/2}\xi^A.\nonumber
\end{align}

The Eulerian variation of the density and pressure at linear order is related to the change of the baryon number density as follows. First, we assume that the baryon number of each fluid element in the perturbed configuration is conserved:
\begin{align}
    \Delta\left[\nabla_\mu \left(n u^\mu\right)\right]=0,\label{eq:baryonnumberconservation}
\end{align}
where $n$ is the baryon number density; $\Delta n$ is its Lagrangian displacement. Using Eq.~\eqref{eq:Deltaumu}, Eq.~\eqref{eq:baryonnumberconservation} at linear order leads to the fractional change of the baryon number density,
\begin{align}
    \frac{\Delta n}{n}=-\frac{1}{2} \left(g^{\mu\nu}+u^\mu u^\nu\right) \Delta g_{\mu \nu}, 
\end{align}
which reads
\begin{align}
    &\frac{\Delta n}{n}=- e^{-i \omega t} Y_{\ell m}\nonumber\\
    & \times \left[\frac{H_2}{2}+K +\ell\left(\ell+1\right)r^{\ell-2}V \right. \nonumber \\
    & \left. +e^{-\lambda/2}r^{\ell-2}\left\{(\ell+1)W+ rW'\right\}\right].
\end{align}

Now, we assume that the perturbation is isentropic~($s={\rm const.}$ everywhere), $\Delta s=0$ for the entropy~$s$ per baryon.\footnote{The isentropic perturbation is a good approximation within the framework of static tides. This condition is equivalent to assuming that perturbative configurations follow a barotropic EoS.  Note that the assumption of isentropic perturbations can be subtle in the presence of out-of-equilibrium corrections, such as viscous effects.} Then, the first law of thermodynamics implies
\begin{align}
    \Delta \rho=\left(\rho+p\right)\frac{\Delta n}{n}.
\end{align}
This is compatible with the baryon conservation, $\nabla_\mu (n u^\mu)=0$, and the projection of $\nabla_\mu T^{\mu \nu}=0$ along the four velocity of the fluid.

The EoS links the pressure and energy density at the background level.
In general, the Lagrangian perturbation to pressure, $\Delta p$, depends on the EoS in equilibrium configurations and on chemical reactions during the dynamical process, which correlates with the entropy per baryon and the change in the fractional abundance of nuclear species inside the NS. If the relaxation timescale of perturbed configurations toward equilibrium is much shorter than the orbital period, the reactions may maintain equilibrium even in perturbed configurations in the adiabatic manner. The fractional abundance of nuclear species then shifts continuously while maintaining equilibrium, and hence, the variation in pressure with respect to that in density in perturbations approximately follows the same relation as in equilibrium configurations.  
In such a situation, the EoS for the background also holds for perturbed configurations. The connection between density and pressure perturbations is provided by $\gamma=\gamma(\rho)$ through 
\begin{align}
    \frac{\Delta p}{p}= \gamma \frac{\Delta n}{n}. 
\end{align}
With these, one can obtain the Eulerian variation of the pressure and energy density via
\begin{align}
    \delta \rho=& \Delta\rho -\xi^r \rho',\quad \delta p= \Delta p-\xi^r p'.
\end{align}

\subsubsection{Perturbation equations}
Now, let us linearize Einstein's equations and the matter conservation law, $\delta G_{\mu \nu}=8\pi \delta T_{\mu \nu}$ and $ \delta (\nabla_\mu T^{\mu \nu})=0$. Combining various components in these equations, we derive a set of perturbation equations for $H,W$, and $V$ as follows. 

We begin with the perturbed Einstein's equations.
First,  $\delta G_\theta^{~\theta}- \delta G_\varphi^{~\varphi}=8\pi (\delta T_\theta^{~\theta}-\delta T_\varphi^{~\varphi})$ allows us to express $H_2$ in terms of $H$. With this, $\delta G_r^{~\theta}=8\pi \delta T_r^{~\theta}$ yields $H_1=H_1(H, K,H',K')$; $\delta G_t^{~\theta}=8\pi \delta T_t^{~\theta}$ leads to $H_1'=H_1'(H,K,V,H',K')$. Eliminating $H_2$, $H_1'$, and $H_1$ from $\delta G_r^{~r}=8\pi \delta T_r^{~r}$ and $\delta G_r^{~t}=8\pi \delta T_r^{~t}$, one obtains $K$ and $K'$ in the form of a combination of $(H, W, V,H',W',V')$. One can then express $H''$ in terms of $(H,W,V,H',W',V')$ using $\delta G_t^{~t}=8\pi \delta T_t^{~t}$. 

Next, we look at the perturbed stress-energy conservation equation.
From $\delta (\nabla_\mu T^\mu_{~\theta}) =0$, one derives the expression for $W'$ in terms of $(H,W,H',V')$, leading to $\delta (\nabla_\mu T^\mu_{~\varphi}) =0$. Next, $\delta (\nabla_\mu T^\mu_{~r})=0$ gives $V'$ in terms of $(H,W,V,H')$, thereby allowing one to eliminate $V'$ from $W'$ and $H''$. 

Thus, the perturbation equations to be solved are summarized into a set of the following three equations in the form of
\begin{align}
    &H''=\alpha_{H,H'}H'+\alpha_{H,H} H+\alpha_{H,W}W+\alpha_{H,V}V, \label{eq:EqforH}
\end{align}
and
\begin{align}
    &W'=\alpha_{W,H'} H'+\alpha_{W,H}H+\alpha_{W,W}W+\alpha_{W,V}V,\label{eq:EqforW}
\end{align}
with
\begin{align}
    &V'=\alpha_{V,H'} H'+\alpha_{V,H}H+\alpha_{V,W}W+\alpha_{V,V}V.\label{eq:EqforV}
\end{align}
Here, $\alpha_{q,i}$ are functions of the background quantities. We provide their expression in online~\cite{git}.

\subsubsection{Static tidal perturbations}
Taking the stationary limit~$\omega\to 0$, Eq.~\eqref{eq:EqforW} leads to an algebraic equation for $W$, allowing one to express it in terms of $H$. Then, Eq.~\eqref{eq:EqforV} connects $V$ with $H$. One can now show $\Delta n/n={\cal O}(\omega)$. In other words, the fractional volumetric change per baryon is zero at the leading order of small~$\omega$ expansion. This leads to $\delta \rho= c_s^2 \delta p$ for the speed of sound in unperturbed configurations,~$c_s:=(dp/d\rho)^{1/2}$, which is the assumption often used in the literature. Equation~\eqref{eq:EqforH} thus results in a single second-order ordinary differential equation for $H$:
\begin{align}
    &H''+P_1 H'-P_2 H=0,\label{eq:mastereqforEven}
\end{align}
with 
\begin{align}
    P_1=& \frac{2}{r}+e^{\lambda}\left[\frac{2m}{r^2}+4\pi r\left(p-\rho\right)\right] ,\label{eq:P1P2}\\
    P_2=&e^\lambda\left[\frac{\ell\left(\ell+1\right)}{r^2}-4\pi \frac{ p+\rho}{c_s^2}-4\pi \left(5\rho+9p\right)\right]  +\nu'^2. \nonumber
\end{align}
Equation~\eqref{eq:mastereqforEven} is consistent with Eq.~(15) of Ref.~\cite{Hinderer:2007mb} and Eq.~(27) of Ref.~\cite{Damour:2009vw}.

The tidal deformabilities,~$\bar{\lambda}_2$ and $\bar{\lambda}_3$, in Figs.~\ref{Fig:Lambda2Sigma2}, \ref{Fig:Lambda2Lambda3}, and~\ref{Fig:ILove} are derived by solving Eq.~\eqref{eq:mastereqforEven} with the background being the modified Tolman~VII solution, provided in Appendix~\ref{Appendix:modifiedTolmanVII}. In what follows, we solve Eq.~\eqref{eq:mastereqforEven} in the stellar exterior and interior on the fully analytic background, provided in Appendix~\ref{Appendix:modifiedTolmanVII}, in an analytic manner. We then match the logarithmic derivatives of the solutions at the stellar surface, thereby describing the static tidal response.

\subsubsection{External problem}
The exterior solution, $H_{\rm ext}$, corresponds to tidal perturbations in the Schwarzschild spacetime~\eqref{eq:Schwarzschildmetric}, and thus obeys
\begin{align}
    &H_{\rm ext}''+\frac{2\left(r-M\right)}{r^2f}H_{\rm ext}'-\frac{1}{r^2 f}\left[\ell\left(\ell+1\right)+\frac{4M^2}{r^2f}\right]H_{\rm ext}=0,
\end{align}
where $f:=1-2M/r$. The general solution is~\cite{Katagiri:2024fpn,Katagiri:2024wbg}
\begin{align}
    H_{\rm ext}=\mathbb{E}_H H_T+\mathbb{I}_H H_R,\label{eq:Hext}
\end{align}
where $\mathbb{E}_{H}$ and $\mathbb{I}_{H}$ are integration constants with
\begin{align}
H_T:=&f\left(\frac{r}{M}\right)^\ell~_2F_1\left(-\ell+2,-\ell;-2\ell;2M/r\right),\\
H_R:=&f\left(\frac{M}{r}\right)^{\ell+1}~_2F_1\left(\ell+1,\ell+3;2\ell+2;2M/r\right).\nonumber
\end{align}
Here, $_2 F_1(a,b;c;2M/r)$ are Gaussian hypergeometric functions~\cite{NIST:DLMF}. Following a ``standard" definition~\cite{Binnington:2009bb,Poisson:2020vap,Katagiri:2024wbg} arising from the analytical properties of the hypergeometric functions, one can relate $\mathbb{E}_{H}$ and $\mathbb{I}_{H}$ to the electric-type tidal moment and mass multipole moment, thereby obtaining the expression for the electric-type TLNs,
\begin{align}
    k_\ell=\frac{1}{2}{\cal C}^{2\ell+1}\frac{\mathbb{I}_H}{\mathbb{E}_H}.\label{eq:kell}
\end{align}
The integration constants are determined by matching the exterior solution with an interior solution, derived in the subsequent section, at the stellar surface.

\subsubsection{Internal problem}
We first introduce a new variable $y_{\rm in}^+ =r H_{\rm in}' /H_{\rm in}$ for the interior solution~$H_{\rm in}$ for later convenience. Equation~\eqref{eq:mastereqforEven} reduces to
\begin{align}
    {ry_{\rm in}^+}' +\left(y_{\rm in}^+\right)^2+F^+ y_{\rm in}^+-Q^+=0,\label{eq:eqforyeven}
\end{align}
with 
\begin{align}
    F^+= & e^\lambda \left[1+4\pi r^2 \left(p-\rho\right)\right],\\
    Q^+=& \ell\left(\ell+1\right)e^\lambda -4\pi r^2\left\{e^\lambda\left[5\rho+9p +\frac{\rho+p}{c_s^2}\right]-\frac{\nu'^2}{4\pi}   \right\}.\nonumber
\end{align}
The regularity condition for $H$ at the stellar origin translates into $y_{\rm in}^+|_{r=0}=\ell$. 

To solve Eq.~\eqref{eq:eqforyeven} analytically, we first use a PM expansion for $y_{\rm in}^+$ in terms of ${\cal C}$, following Refs.~\cite{Chan:2014tva,Jiang:2020uvb,Lowrey:2024anh},
\begin{align}
    y_{\rm in}^+=\sum_{j=0}^{j_{\rm max}^+}{\cal C}^j y_{j}^+\left(r\right).\label{eq:PMexpansionofyinEven}
\end{align}
Equation~\eqref{eq:eqforyeven} is then reduced into a set of equations for $y_{j}^+$. We further Taylor-expand $y_{j}^+$ around the origin,
\begin{align}
    y_j^+=\sum_{k=0}^{k_{\rm max}} c_{j,2k}^+  \xi^{2k}.\label{eq:TaylorExpofyEven}
\end{align}
The resulting equations form recurrence relations about $c_{j,2k}^+$. Solving them under $y_{\rm in}^+|_{r=0}=\ell$, we derive the analytic expression for $y_{\rm in}^+$, thereby obtaining $y_{\rm in}^+|_{r=R}$. Note that $(j_{\rm max}^+,k_{\rm max})$ is not specified at the current stage; the best set is determined by comparing analytically computed~$\bar{\lambda}_\ell$ with numerical results for various sets.

\subsubsection{Electric-type tidal deformability}
The following matching, $r H_{\rm ext}'/H_{\rm ext}|_{r=R}=y_{\rm in}^+|_{r=R}$, allows us to obtain $k_\ell$ in Eq.~\eqref{eq:kell} in the form of the expansion about ${\cal C}$ via
\begin{align}
    \left.\frac{\mathbb{I}_H}{\mathbb{E}_H}=-\frac{RH_T'-y_{\rm in}^+ H_T}{RH_R'-y_{\rm in}^+ H_R}\right|_{r=R}.
\end{align}
Note that $\mathbb{I}_H/\mathbb{E}_H$ scales as ${\cal O}({\cal C}^{-2\ell-1})$, implying $k_\ell={\cal O}({\cal C}^0)$. This is consistent with previous findings by Refs.~\cite{Damour:2009vw,Binnington:2009bb} showing that the electric-type TLNs for stars are generically nonvanishing in the Newtonian limit. 

We thus obtain the expression for $\bar{\lambda}_\ell$ in Eq.~\eqref{eq:electriclambda} in the form of a polynomial of ${\cal C}$. To improve its accuracy further, we re-sum the polynomial with the Pad{\'e} approximant, thereby obtaining
\begin{align}
    \bar{\lambda}_\ell= \frac{2}{\left(2\ell-1\right)!!} \frac{1}{{\cal C}^{2\ell+1}}\frac{p_{\ell,0}^+ + p_{\ell,1}^+ {\cal C} + p_{\ell,2}^+ {\cal C}^2  + p_{\ell,3}^+ {\cal C}^3  }{ q_{\ell,0}^+ + q_{\ell,1}^+ {\cal C} + q_{\ell,2}^+ {\cal C}^2 +q_{\ell,3}^+ {\cal C}^3 },\label{eq:barlambdainCinApp}
\end{align}
where $p_{\ell,i}^+=p_{\ell,i}^+(\alpha)$ and $q_{\ell,i}^+=q_{\ell,i}^+(\alpha)$ (with the modified Tolman $\alpha$) for $i=0,1,2,3$. The explicit forms of $p_{\ell,i}^+$ and $q_{\ell,i}^+$ are provided in online~\cite{git}. Equation~\eqref{eq:barlambdainCinApp} is identical to Eq.~\eqref{eq:barlambdainC}.

\begin{figure}[htbp]
\centering
\includegraphics[scale=0.68]{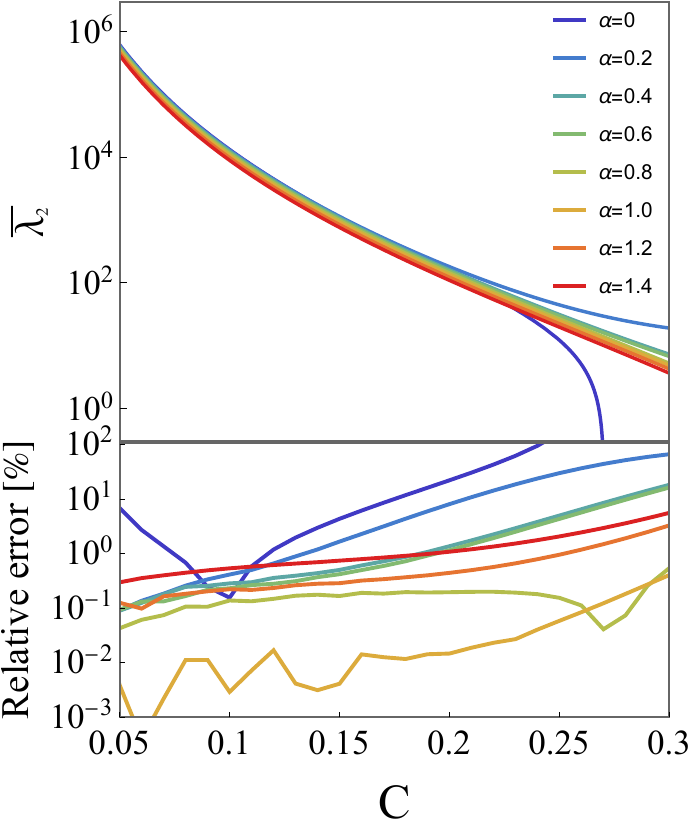}
\hspace{0.3cm}
\caption{(Top panel)~The electric-type quadrupolar tidal deformability~$\bar{\lambda}_2$ computed by Eq.~\eqref{eq:barlambdainCinApp}.  Note that the PM expansion is truncated at $(j_{\rm max},k_{\rm max})=(6,15)$ in Eq.~\eqref{eq:PMexpansionofyinEven}. (Bottom panel)~ The relative error between their values and numerical results on the modified Tolman~VII background for various~$\alpha$.}
\label{Fig:Lambda2num}
\end{figure}

Figure~\ref{Fig:Lambda2num} compares $\bar{\lambda}_2$ computed from Eq.~\eqref{eq:barlambdainCinApp} with the values computed numerically, showing good agreement across various~$\alpha$ in the low-compactness regime. The agreement between the analytical and numerical solutions for $\bar{\lambda}_{2}$ and $\bar{\lambda}_{3}$ follows the same trend. In the regime of low compactness, the values found analytically exhibit excellent agreement with the numerical results, with the relative error decreasing until reaching a minimum at a certain value of compactness. Beyond this point, the error begins to increase again and eventually diverges rapidly as ${\cal C}$ increases. The main effect of varying the truncation order of ${\cal C}$ seems to shift the location of this minimum: a higher truncation order of ${\cal C}$ moves this minimum towards a higher value of compactness. This, in turn, extends the regime over which the analytic expressions fo $\bar{\lambda}_{2}$ or $\bar{\lambda}_{3}$ remain accurate. One can further reduce the error by employing higher order Pad{\'e} approximants. Nevertheless, for the purposes of the present work, it is sufficient to consider the Pad{\'e} approximants up to $[3/3]$ order. Truncating the PM expansion at ${\cal O}({\cal C}^6)$ is sufficient for the Pad{\'e} approximation at $[3/3]$~order, as proceeding to even higher orders in ${\cal C}$ yields only negligible improvements in the accuracy of both $\bar{\lambda}_{2}$ and $\bar{\lambda}_{3}$.

Furthermore, we compared our results with the analytical expression derived in Ref.~\cite{Jiang:2019vmf} for $\bar{\lambda}_{2}$. We find that our expression yields a more accurate $\bar{\lambda}_{2}$ over the entire range of compactness by expanding $y_{\rm in}^+$ to higher order about ${\cal C}$ and $\xi$, thereby improving the previous results.

\subsubsection{Compactness in terms of $\bar{\lambda}_2$}
Here, we provide the expressions for $\beta_j$ in Eq.~\eqref{eq:Cinbarlambda}:
\begin{align}
\beta_1=& 0.735783 \dots,\nonumber\\
\beta_2=& -0.692724 \dots ,\nonumber\\
\beta_3=&0.502849 \dots ,\nonumber\\
\beta_4=&-0.313895 \dots ,\label{eq:CoefficientsofC}\\
\beta_5=& 0.164826 \dots,\nonumber\\
\beta_6=& -0.0694865 \dots,\nonumber\\
\beta_7=&  0.0223008 \dots,\nonumber\\
\beta_8=& -0.00777183 \dots.\nonumber
\end{align}
Here, $\dots$ denotes higher-order decimal digits. In the actual derivation of the power-law expressions~\eqref{eq:lambda2sigma2} and~\eqref{eq:lambda2lambda3}, we have used fractions, which have arbitrary precision. We only present numerical values above because the exact expressions are too lengthy. The precise values of $\beta_j$ are provided in online~\cite{git}.

\subsection{Odd-parity sector}

\subsubsection{Stellar perturbation equations}
We next present stellar perturbations in the odd-parity sector.
In the Regge-Wheeler gauge, the odd-parity sector of metric perturbations with harmonic decomposition takes the form,
\begin{align}
     &h_{\mu \nu}^{-}dx^\mu dx^\nu=\left(2 h^{\ell m} S_\theta^{\ell m} dt d\theta+2 h^{\ell m} S_\varphi^{\ell m} dt d\varphi\right.\nonumber \\
    &\left.+2 h_1^{\ell m} S_\theta^{\ell m} dr d\theta+2 h_1^{\ell m} S_\varphi^{\ell m} dr d\varphi\right) e^{-i \omega t}\label{eq:hodd}.
\end{align}
Here, $h^{\ell m}=h^{\ell m}(r)$, $h_1^{\ell m}=h_1^{\ell m}(r)$; $(S^{\ell m}_\theta,S^{\ell m}_\varphi):=(-\partial_\varphi Y_{\ell m}/\sin \theta, \sin \theta \partial_\theta Y_{\ell m})$. Henceforth, we drop the superscript~$\ell m$. 

In the odd-parity sector, the metric perturbation is not coupled to the pressure and density perturbations but induces the displacement of the four velocity. The Eulerian perturbation of the four velocity is parametrized by a function~$U=U(r)$~\cite{1969ApJ...155..163P,Pani:2015nua,Pani:2018inf}
\begin{align}
    \delta u^t=& 0,\quad    \delta u^r= 0,\\
    \delta u^A = & \frac{U  e^{\nu /2 }}{4\pi r^2 \left(\rho+p\right)}\left(S_\theta^{\ell m},\frac{S_\varphi^{\ell m}}{\sin^2\theta}\right)e^{-i \omega t}.
\end{align}

Now, linearizing Einstein's equations and the matter conservation law, we derive a set of perturbation equations. First, $\delta G_\theta^{~\theta}=8\pi \delta T_\theta^{~\theta}$ allows us to express $h_1'$ in terms of $h$ and $h_1$, Then, 
$\delta G_r^{~\theta}=8\pi \delta T_r^{~\theta}$ leads to 
\begin{align}
h_1 =& i \omega r\frac{r h' - 2h}{\omega^2 r^2 -e^{\nu} \left(\ell^2+\ell-2\right)}.
\end{align}
One finds from $\delta (\nabla_\mu T^{\mu}_{~\theta})=0$ that
\begin{align}
U =& - 4 \pi e^{-\nu} h (\rho+p).
\end{align}
With these, $\delta G_t^{~\theta}=8\pi \delta T_t^{~\theta}$ yields a second-order differential equation for $h$:
\begin{align}
h''= \alpha_{h,h'} h' + \alpha_{h,h}h, \label{eq:eqforh0}
\end{align}
where $\alpha_{h,i}$ are functions of the background quantities.  We provide their expression in online~\cite{git}.

\subsubsection{Static tidal perturbation}
Taking the stationary limit of Eq.~\eqref{eq:eqforh0} leads to\footnote{A master equation for stationary perturbations in the odd-parity sector depends on the assumption for the fluid state~\cite{Landry:2015cva}. In particular, the stationary limit of frequency-dependent perturbations forces the fluid be an irrotational state that permits internal motions~($U|_{\omega \to 0}\neq 0$), whereas strictly static perturbations require strict hydrostatic equilibrium, even in the perturbed configuration~($U=0$ for $\omega=0$)~\cite{Pani:2018inf}. The former state is more realistic compared to the latter. Equation~\eqref{eq:mastereqforOdd} assumes the irrotational state, in which the vorticity vector vanishes~\cite{Pani:2018inf}.}
\begin{align}
    &h''-4\pi r e^\lambda\left(\rho+p\right)h'\nonumber\\
    &-e^\lambda\left[\frac{\ell\left(\ell+1\right)}{r^2}-\frac{4m}{r^3}-8\pi \left(\rho+p\right)\right]h=0 \label{eq:mastereqforOdd}.
\end{align}
Equation~\eqref{eq:mastereqforOdd} is consistent with the stationary limit of Eq.~(11) in Ref.~\cite{Pani:2018inf}, Eq.~(5.6) in Ref.~\cite{Landry:2015cva}, and Eq.~(31) in Ref.~\cite{Damour:2009vw}.

The tidal deformability~$\bar{\sigma}_2$ is derived by solving Eq.~\eqref{eq:mastereqforOdd} numerically under the modified Tolman~VII background solution. Here, similar to the even-parity case, we solve Eq.~\eqref{eq:mastereqforOdd} in the stellar exterior and interior on the analytic background in a fully analytic manner and then match their logarithmic derivatives of the solutions at the stellar surface, thereby evaluating the static tidal response.

\subsubsection{External problem}
Equation~\eqref{eq:mastereqforOdd} reduces to a master equation for tidal perturbations, $h_{\rm ext}$, in the Schwarzschild spacetime~\eqref{eq:Schwarzschildmetric},
\begin{align}
    h_{\rm ext}''-\frac{1}{r^2f}\left[\ell\left(\ell+1\right)-\frac{4M}{r}\right]h_{\rm ext}=0.
\end{align}
The general solution is found in closed form~\cite{Katagiri:2024fpn,Katagiri:2024wbg}
\begin{align}
   h_{\rm ext}=\mathbb{E}_h h_T+\mathbb{I}_h h_R,\label{eq:hinvacuum}
\end{align}
where $\mathbb{E}_{h}$ and $\mathbb{I}_{h}$ are integration constants with
\begin{align}
h_T:=&\left(\frac{r}{M}\right)^{\ell+1}~_2F_1\left(-\ell+1,-\ell-2;-2\ell;2M/r\right),\\
h_R:=&\left(\frac{M}{r}\right)^{\ell}~_2F_1\left(\ell-1,\ell+2;2\ell+2;2M/r\right).\nonumber
\end{align}
Following a ``standard" definition~\cite{Binnington:2009bb,Poisson:2020vap,Katagiri:2024wbg}, one can identify the integration constants with the magnetic-type tidal moment and the current multipole moment, thus obtaining
\begin{align}
    j_\ell=-\frac{\ell+2}{\ell-1}{\cal C}^{2\ell+1}\frac{\mathbb{I}_h}{\mathbb{E}_h}.\label{eq:jell}
\end{align}
Similar to the even-sector case, the integration constants are determined by matching the exterior solution with an interior solution derived in the next section. 

\subsubsection{Internal problem}
Introducing a new variable~$y_{\rm in}^- =rh_{\rm in}'/h_{\rm in}$ for convenience, Eq.~\eqref{eq:mastereqforOdd} is cast into
\begin{align}
    r {y_{\rm in}^-}'+\left(y_{\rm in}^-\right)^2+F^- y_{\rm in}^- -Q^-=0,\label{eq:eqforyodd}
\end{align}
with 
\begin{align}
    F^-=& -1-4\pi r^2 e^\lambda \left(\rho+p\right),\\
    Q^-=& -e^\lambda \left\{ \frac{4m}{r}  -\left[ \ell\left(\ell+1\right)  -8\pi r^2 \left(\rho+p\right) \right]   \right\}.\nonumber
\end{align}
The regularity condition for $h$ at the stellar origin translates to $y_{\rm in}^-|_{r=0}=\ell+1$.

Similar to the even-parity case, we expand $y_{\rm in}^-$ in terms of ${\cal C}$,
\begin{align}
    y_{\rm in}^- =\sum_{j=0}^{j_{\rm max}^-} {\cal C}^j y_j^-(r).\label{eq:PMexpansionofyinOdd}
\end{align}
Substituting this into Eq.~\eqref{eq:eqforyodd}, we derive differential equations for $y_j^-$. We solve them order by order and impose $y_{\rm in}^-|_{r=0}=\ell+1$, thereby deriving  $y_{\rm in}^-|_{r=R}$ approximately. Notably, the construction of $y_{\rm in}^-$ does not require a Taylor expansion around the origin, unlike to the even-parity case as in Eq.~\eqref{eq:TaylorExpofyEven}.

\subsubsection{Magnetic-type tidal deformability}
\begin{figure}[htbp]
\centering
\includegraphics[scale=0.75]{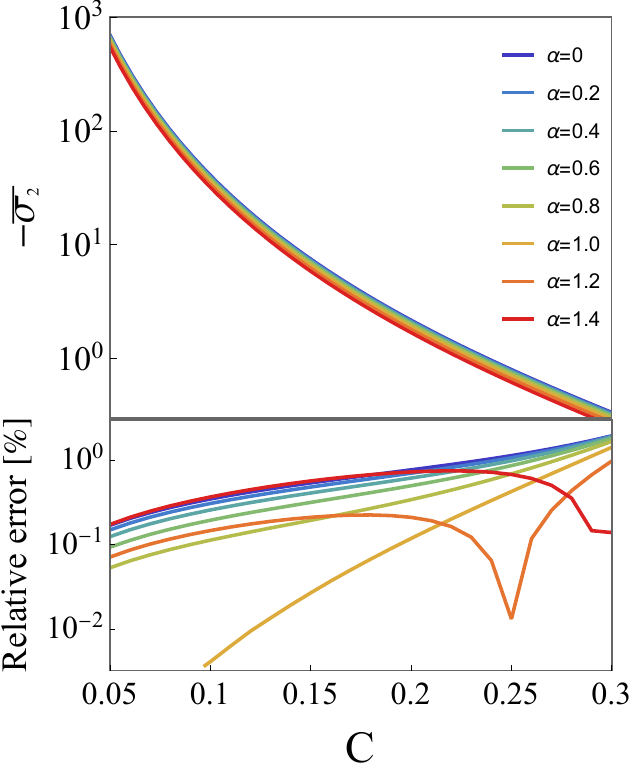}
\hspace{0.3cm}
\caption{(Top panel)~The magnetic-type quadrupolar tidal deformability~$\bar{\sigma}_2$ computed by Eq.~\eqref{eq:barsigmainCinApp}.  Note that the PM expansion is truncated at $j_{\rm max}=5$ in Eq.~\eqref{eq:PMexpansionofyinOdd}. (Bottom panel)~ The relative error between their values and numerical results on the modified Tolman~VII background for various~$\alpha$.}
\label{Fig:Sigma2num}
\end{figure}

The matching condition, $r h_{\rm int}'/h_{\rm int}|_{r=R}=y_{\rm ext}^-|_{r=R}$, leads to the expression for $j_\ell$ in Eq.~\eqref{eq:jell} in the form of the expansion of ${\cal C}$ through
\begin{align}
    \left.\frac{\mathbb{I}_h}{\mathbb{E}_h}=-\frac{Rh_T'-y_{\rm in}^- h_T}{Rh_R'-y_{\rm in}^- h_R}\right|_{r=R}.
\end{align}
Note that $\mathbb{I}_h/\mathbb{E}_h$ scales as ${\cal O}({\cal C}^{-2\ell})$, implying $j_\ell={\cal O}({\cal C}^1)$. The magnetic-type TLNs have no Newtonian analogue, as shown in Refs.~\cite{Damour:2009vw,Binnington:2009bb}.

We thus obtain $\bar{\sigma}_\ell$ in Eq.~\eqref{eq:magneticsigma} in the form of a polynomial of ${\cal C}$. The Pad{\'e} approximant of the polynomial provides more accurate expressions,
\begin{align}
    \bar{\sigma}_\ell =  \frac{\ell-1}{4\left(\ell+2\right)\left(2\ell-1\right)!!} \frac{1}{{\cal C}^{2\ell}}\frac{p_{\ell,0}^- + p_{\ell,1}^- {\cal C} + p_{\ell,2}^- {\cal C}^2  }{ q_{\ell,0}^- + q_{\ell,1}^- {\cal C} + q_{\ell,2}^- {\cal C}^2 }, \label{eq:barsigmainCinApp}
\end{align}
where $p_{\ell,i}^-=p_{\ell,i}^-(\alpha)$ and $q_{\ell,i}=q_{\ell,i}(\alpha)$ for $i=0,1,2$. Equation~\eqref{eq:barsigmainCinApp} is identical to Eq.~\eqref{eq:barsigmainC}. The explicit forms of $p_{\ell,i}^-$ and $q_{\ell,i}^-$ are provided in online~\cite{git}.

Figure~\ref{Fig:Sigma2num} compares $\bar{\sigma}_2$ computed from Eq.~\eqref{eq:barsigmainCinApp} with the values computed numerically, showing their excellent agreement across various~$\alpha$.

\section{I-Love relation from low compressibility}\label{Appendix:ILove}
\begin{figure}[htbp]
\centering
\includegraphics[scale=0.70]{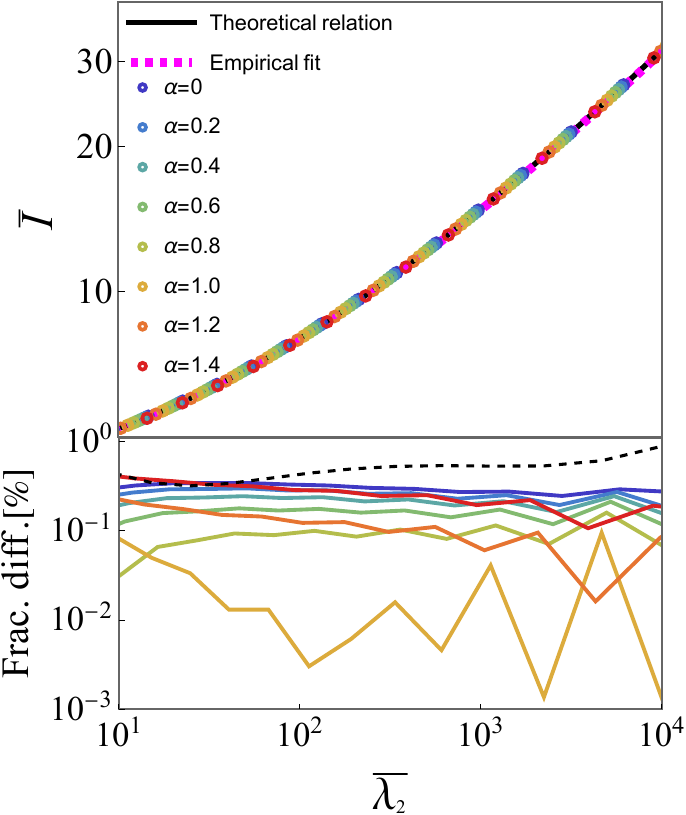}
\hspace{0.3cm}
\caption{Similar to Fig.~\ref{Fig:Lambda2Sigma2} but for the I-Love relation.}
\label{Fig:ILove}
\end{figure}

The moment of inertia quantifies how rapid an object can rotate for a given angular momentum and is related to the current dipole moment of the object. One can extract the (normalized) moment of inertia,~$\bar{I}$, from the asymptotic expansion of the~$t\varphi$ component of the metric of spinning stars at large distances. In this work, we utilize the analytic expression for $\bar{I}$ of the fully analytic modified Tolman~VII model, derived in Ref.~\cite{Jiang:2020uvb}. The expression for $\bar{I}$ can be found in Ref.~\cite{onlinelinkforI}. Substituting ${\cal C}$ in Eq.~\eqref{eq:Cinbarlambda} into this expression with $\alpha=1$, and expanding it about $1/\bar{\lambda}_2^{1/5}$, we obtain the theoretical power-law I-Love relation,
\begin{widetext}
\begin{align}
    \bar{I}= 0.527755 \bar{\lambda}_2^{2/5}\left(1+\frac{2.71536}{\bar{\lambda}_2^{1/5}}  +\frac{2.95181}{\bar{\lambda}_2^{2/5}}  +\frac{1.51947}{\bar{\lambda}_2^{3/5}} +\frac{0.298088}{\bar{\lambda}_2^{4/5}} +\frac{0.0295906}{\bar{\lambda}_2} \right). \label{eq:lambda2I}
\end{align}
\end{widetext}
Figure~\ref{Fig:ILove} presents the I-Love relation in the modified Tolman~VII solution, showing that the power-law relation~\eqref{eq:lambda2I} agrees well with the empirical fit in Refs.~\cite{Yagi:2013bca,Yagi:2013awa}. 

We discuss the underlying structure of the I-Love relation. Let us focus on the Newtonian-order contribution. Noting that $\bar{I}$ scales as ${\cal O}({\cal C}^{-2})$, one can obtain
\begin{align}
    \bar{I} \propto \frac{\bar{I}^{\rm(N)}}{\left(k_2^{\rm (N)} \right)^{2/5}} \bar{\lambda}_2^{2/5},\label{eq:ILoveatleading}
\end{align}
where $\bar{I}^{\rm(N)}$ is the leading-order coefficient of the PM expansion of $\bar{I}$~\cite{Jiang:2020uvb}. Here, notice that the power-law exponent of $\bar{\lambda}_2$, $2/5$, is independent of any microphysics, and is determined by the scaling law of $\bar{I}$ and $\bar{\lambda}_2$ with respect to ${\cal C}$, as in Eq.~\eqref{eq:EvenOddatleading}. Now, similar to Eq.~\eqref{eq:k2expansionaroundalpha0}, we expand $\bar{I}^{\rm (N)}$ around $\alpha=\alpha_0$, obtaining
\begin{align}
    \bar{I}^{\rm (N)}= C_{I}^{(0)}\left[1- C_{I}^{(1)} \left(\alpha-\alpha_0\right)+{\cal O}\left(\left(\alpha-\alpha_0\right)\right)^2 \right],\label{eq:barIexpansion}
\end{align}
where $C_{I}^{(i)}$ are positive numbers. For example, $( C_{I}^{(0)}, C_{I}^{(1)})=(0.29,0.11)$ for $\alpha_0=1$. Notice that the ratio of the coefficient at first order in the expansion to that at zeroth order is negative as the expansion of $k_2$ is in Eq.~\eqref{eq:k2expansionaroundalpha0}. Physically, for stiffer EoSs, the mass distribution in the stellar interior peaks further outward, making rotation more difficult for a given angular momentum. Also note $0.05 \lesssim C_I^{(1)} \lesssim 0.16$ and $C_{I}^{(1)}\simeq 0.35 C_{k_2}^{(1)}$ for $0 \le \alpha_0 \le 1.4$, indicating the modest~$\alpha$ dependence. Indeed, there exists the so-called I-C relation~\cite{Yagi:2016bkt}, which is assessed by 
\begin{align}
\frac{\left.\bar{I}^{\rm (N)}\right|_{\alpha=0} }{\left.\bar{I}^{\rm (N)}\right|_{\alpha=1.4}} \simeq 1.1.\label{eq:RatioofI}
\end{align}
For comparison, the ratio for polytropes with $0\le n \le 1$ is~\cite{Yagi:2013bca,Yagi:2013awa}
\begin{align}
    \frac{\left.\bar{I}^{\rm (N)}\right|_{n=0} }{\left.\bar{I}^{\rm (N)}\right|_{n=1}} \simeq 1.5.
\end{align}
The I-C relation has weaker universality compared to the I-Love relation~($\sim 0.2\%$ difference between polytropes with~$0\le n \le 1$)~\cite{Yagi:2013bca,Yagi:2013awa}. The ratio,
\begin{equation}
   r_{\bar \lambda_2-\bar I}^{(1)}(\alpha) \equiv \frac{C_I^{(1)} (\alpha)}{C_{k_2}^{(1)}(\alpha)}\simeq 0.35
\end{equation} 
approximately holds for the range of  $\alpha$ studied in this paper and is close to the power-law exponent of the I-Love relation,~$2/5$, as presented in Table~\ref{Table:ratioforILove}. As $\alpha$ further increases, the ratio deviates from this value.
\begin{table}[t]
\begin{tabular}{ |c|c|c|c| } 
\hline
 &  $r_{\bar \lambda_2-\bar I}^{(1)}(1)$ 
 & Exponent $\bar n$ & Ratio $r_{\bar \lambda_2-\bar I}^{(1)}(0)/r_{\bar \lambda_2-\bar I}^{(1)}(1.4)$ \\
\hline
$\bar{\lambda}_2- \bar{I}$ & 0.35 & 2/5 & 1.1 \\ 
\hline
\end{tabular}
\caption{Similar to Table~\ref{Table:ratio} but for the I-Love relation.}
\label{Table:ratioforILove}
\end{table}

Equation~\eqref{eq:ILoveatleading} shows that $k_2$ enters in the denominator of the coefficient of the power-law relation in the same way as in Eq.~\eqref{eq:EvenOddatleading}. The $\alpha$ dependence coming from $k_2^{\rm (N)}$ contributes to the coefficient in the opposite way to that from $\bar{I}^{\rm (N)}$ in the combination. One finds
\begin{align}
    \frac{\bar{I}^{\rm(N)}}{\left(k_2^{\rm (N)} \right)^{2/5}} \propto  1-\left(C_{I}^{(1)}-\frac{2}{5}C_{k_2}^{(1)}\right)\left(\alpha-\alpha_0\right)  +{\cal O}\left(\left(\alpha-\alpha_0\right)^2 \right).\label{eq:powerlawcoefficientatleadingPMforILove}
\end{align}
Recalling that $C_I^{(1)}\simeq 0.35 C_{k_2}^{(1)}$ and $C_{k_2}={\cal O}(10^{-1})$, the factor in front of $(\alpha-\alpha_0)$ is of ${\cal O} (10^{-2})$, indicating approximate universality. This is the same mechanism as the universal Love relations studied in Section~\ref{Section:PotentialOrigin}. A key observation is that $\bar{I}$ shares the $\alpha$ dependence with $\bar{\lambda}_2$.
\begin{figure}[htbp]
\centering
\includegraphics[scale=0.64]{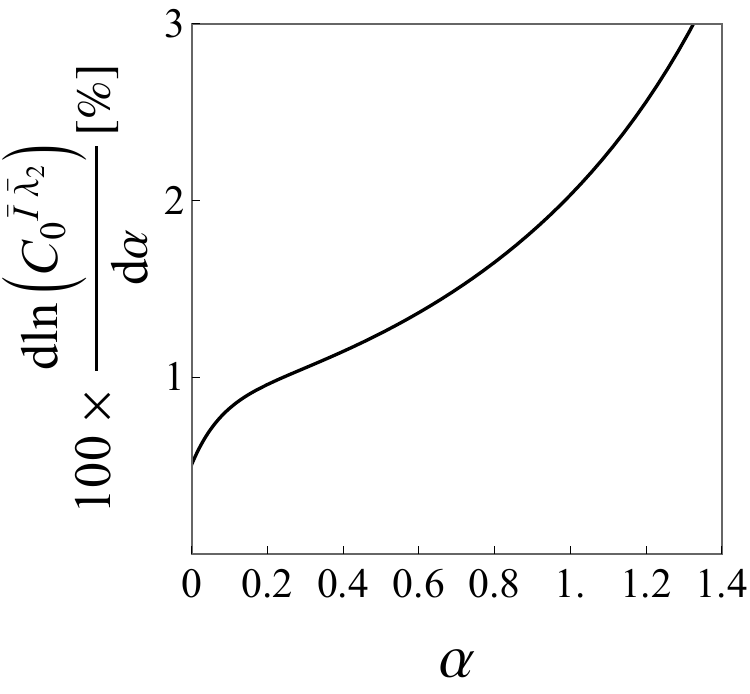}
\hspace{0.3cm}
\caption{The relative change rate of $C_{0}^{\bar{I}\bar{\lambda}_2}$ in Eq.~\eqref{eq:RatioinINk2N} with respect to $\alpha$, indicating approximate universality between $\bar{I}$ and $\bar{\lambda}_2$.}
\label{Fig:dCIlambda}
\end{figure}

We assess the universality by
\begin{align}
    \frac{\left.C_{0}^{\bar{I}\bar{\lambda}_2} \right|_{\alpha=0}}{\left. C_{0}^{\bar{I}\bar{\lambda}_2} \right|_{\alpha = 1.4}}\simeq 0.98,\quad  C_{0}^{\bar{I}\bar{\lambda}_2}:= \frac{{\bar{I}^{\rm(N)}}}{{\left(k_2^{\rm (N)} \right)^{2/5}}}.\label{eq:RatioinINk2N}
\end{align}
This is slightly larger than the one for polytropes with $0\le n \le 1$, which is $0.2\%$~\cite{Yagi:2013bca,Yagi:2013awa}. In contrast to the I-C relation, the polytropes follow the I-Love relation more robustly compared to the modified Tolman~VII solution with $0\le \alpha \le 1.4$. This implies that the I-C relation does not necessarily underlie the I-Love relation, but is rather mutually correlated with the I-Love relation in a non-hierarchical manner. The suppression mechanism of the EoS dependence in the latter arises from different structures as the I-C relation, which appears to originate from an approximate quadratic energy density profile~\cite{Jiang:2020uvb}. We speculate the similarity of the EoS dependence of TLNs and the moment of inertia~($C_I^{(1)}/C_{k_2}^{(1)}\simeq 0.35$ in the current case) plays a certain role in the degree of the I-Love relation.   

Figure~\ref{Fig:dCIlambda} presents the relative change rate of $C_{0}^{\bar{I}\bar{\lambda}_2}$ in Eq.~\eqref{eq:RatioinINk2N} with respect to $\alpha$, showing that the~$\alpha$ dependence of the power-law coefficient is weak, becomes weaker as $\alpha$ decreases, and reaches a minimum at $\alpha=0$. This implies that the I-Love relation can also be attributed to low compressibility in the entire interior region.

\bibliography{Refs}

\end{document}